\documentclass{jfm}

\usepackage{graphicx}
\usepackage{newtxtext}
\usepackage{newtxmath}
\usepackage{natbib}
\usepackage{hyperref}
\hypersetup{
    colorlinks = true,
    urlcolor   = blue,
    citecolor  = black,
}

\newcommand{\RomanNumeralCaps}[1]

\usepackage{xcolor}

\makeatletter
\let\@underjournal\@empty
\let\@journalname\@empty
\let\@j@urnal\@empty

\def\ps@titlepage{%
  \let\@mkboth\@gobbletwo
  \def\@oddhead{}%
  \def\@evenhead{}%
  \def\@oddfoot{\hfil\thepage\hfil}%
  \let\@evenfoot\@oddfoot
}
\@ifundefined{jfm@upperbanner}{}{\let\jfm@upperbanner\@empty}
\@ifundefined{jfm@lowerbanner}{}{\let\jfm@lowerbanner\@empty}
\@ifundefined{jfm@abstractwarning}{}{\let\jfm@abstractwarning\@empty}
\@ifundefined{jfm@pagelimitwarning}{}{\let\jfm@pagelimitwarning\@empty}

\@ifundefined{linenumbers}{}{%
  
  \@ifundefined{nolinenumbers}{}{\nolinenumbers}%
}
\makeatother
\title{Erosion induced by a disk translating toward or away from a granular bed}

\author{Joanne Steiner\aff{1}, Philippe Gondret\aff{1}, Alban Sauret\aff{2,3} \and Cyprien Morize\aff{1}}

\affiliation{
\aff{1}Université Paris-Saclay, CNRS, Laboratoire FAST, F-91405 Orsay, France
\aff{2}University of Maryland, College Park, Department of Mechanical Engineering, MD 20742, USA
\aff{3}University of Maryland, College Park, Department of Chemical and Biomolecular Engineering, MD 20742, USA
}

\begin{document}
\maketitle

\begin{abstract}
Unsteady flows generated when a body approaches or departs from a granular bed arise in swimming, burrowing, and maneuvering devices. Yet, the threshold for grain motion in such transients remains poorly modeled due to the complexity of the flow. In this study, we report laboratory measurements of the onset of erosion when a rigid circular disk is subjected to a single vertical stroke through quiescent water above a granular bed. The stroke length and travel time were varied independently to determine the critical velocity at which the granular bed is eroded for different minimum distances from the bed. Two erosion mechanisms are observed for disk motion towards the bed: during the stroke, the outward squeezing flow erodes grains near the edge, while after stoppage, the starting vortex or associated secondary vortices impinge on the surface. For motion away from the bed, only the early interaction between the inward suction flow and the nascent vortex entrains grains. The resulting dimensionless thresholds clarify the respective roles of radial flows and vortices in transient, impulsively driven erosion.
\end{abstract}

\begin{keywords}
Sediment transport; Immersed granular material; Vortex dynamics 
\end{keywords}

%%%%%%%%%%%%%%%%%%%%%%%%%%%%%%%%%%%%%%%%%%%%%%%%%%%%%%%%%%%%%%
%%%%%%%%%%%%%%%%%%%%% INTRODUCTION %%%%%%%%%%%%%%%%%%%%%%%%%%%
%%%%%%%%%%%%%%%%%%%%%%%%%%%%%%%%%%%%%%%%%%%%%%%%%%%%%%%%%%%%%%

\section{Introduction}
\label{sec:intro}

The erosion and transport of immersed granular materials subject to a fluid flow is a fundamental and complex problem in hydraulics, coastal engineering, and geomorphology. About 150 years ago, the French engineer Du Boys \citep{Duboys1879} was the first to present a rational approach to sediment transport in gravelbed rivers, as reported by \cite{Hager2005}. In steady or statistically steady flows, the threshold is now usually characterized by a critical Shields number, ${\rm Sh}_{c}$, which describes the ratio of the drag force of the fluid to the apparent weight of the particles \cite[][]{Shields1936}. Most studies investigating the ${\rm Sh}_{c}$ values have been done in laboratory studies in situations of uniform shear and pressure-driven currents \cite[][]{White1940,Miller1977,Buffington1997,Loiseleux2005,Ouriemi2007}. These studies show that ${\rm Sh}_{c}$ varies with the flow regime, ranging from ${\rm Sh}_{c} \simeq 0.13 \pm 0.01$ in the laminar regime to typically 0.04 in the turbulent regime. Various field observations have also validated Shields-type criteria for rivers, estuaries, and desert dunes \cite[][]{Iversen1976,Garcia1999,Charru2013}, and further laboratory investigations have highlighted the onset of bedform initiation under turbulent flows \cite[][]{FranklinCharru2011,Franklin2015}. Beyond these threshold estimates, continuum and two-phase flow models have also been developed to capture sediment transport and bedform dynamics under laminar or turbulent shear \citep{ChauchatMedale2010,CharruBouteloupBonomettiLacaze2016,Chauchat2018sheetflow}. However, recent experiments show that the transition to sediment transport is “blurred” as an effect of the turbulent fluctuations, with the density of grains transported downstream that does not vanish sharply at a well-defined ${\rm Sh}_{c}$ \citep{Salevan2017}. Instead, a crossover region is found somewhere below at ${\rm Sh}_{c} \simeq 0.02$ as also reported by \cite{Lajeunesse2010}. 

Many natural and industrial flows are impulsive or highly unsteady, such as burst-like gusts, breaking waves, flow induced by a propeller, and the vortical wake of swimming organisms. In such situations, peak shear stresses arise on the granular bed from local short-lived jets and vortices, which are not captured by the classical steady-flow threshold.
Unsteady, vortex-driven erosion is important in different situations.  For instance, burrowing bivalves and razor clams fluidize sand by pulsed suction flows \cite[][]{Winter2012} and batoid fishes shed starting vortices that excavate pits for feeding or cover \cite[][]{WilliamsonRoshko1988}. Another example is the scouring processes around obstacles such as bridge piers or offshore wind turbine masts, where the horseshoe vortices act at the foot of the structures and wake vortices act downstream, underscoring the coexistence of distinct scour modes governed by local transient vortex structures \cite[][]{Roulund2005,Manes2015,Auzerais2016,Lachaussee2018}.

Local erosion and sediment resuspension by isolated vortices and transient jets have been considered in different studies. Impinging turbulent jets on a granular bed produce a local erosion once their centerline velocity is larger than a critical value \cite[][]{Badr2016,Brunier2017,Sharma2022}, and this flow configuration is used to characterize the soil stability in the so-called Jet Erosion Test \citep{Hanson2004}. The critical Shields number based on the local velocity at the bed level deduced from the self-similar decay of free jets has been shown by \cite{Badr2014} to be  ${\rm Sh}_{c} \simeq 1$ at the onset of erosion for a particle Reynolds number in the range $1 < {\rm Re}_p < 10^2$. When a vortex ring hits a granular bed, it can also erode grains and create craters, as reported by \cite{Munro2009}, \cite{bethke2012}, \cite{Masuda2012}, and \cite{Yoshida2014}. \cite{Munro2009} have found that the critical Shields number based on the impinging velocity of the vortex ring is ${\rm Sh}_{c} \simeq 1.5$ when the particle Reynolds number based on the typical particle settling velocity is ${\rm Re}_p \gtrsim 8$ and ${\rm Sh}_{c} \simeq 7 \, {\rm Re}_p^{-1/2}$ when ${\rm Re}_p \lesssim 8$.

An important flow configuration is encountered in footstep situations \cite[][]{zhang_particle_2008,Kubota2013} where there is a rapid radial outflow or inflow in the thin gap between the body and the bed, and a pressure impulse and azimuthal shear of one or more vortex rings. Human foot motions, such as walking and foot tapping, resuspend particulate matter on the floor and redistribute it, thereby increasing the particle concentration in the air and affecting indoor air quality.

An example of such a flow configuration is the resuspension of dust due to the approach and collision of an object onto a dusty wall \cite[][]{Madler1999,Eames2000}. For the case of a sphere that is suddenly arrested at the wall after a normal approach to the wall at a constant velocity, \cite{Eames2000} have shown that the critical Shields number based on the sphere approach velocity is  ${\rm Sh}_{c} \simeq 3$ when the particle Reynolds number based on the typical particle settling velocity is ${\rm Re}_p \gtrsim 2$ and ${\rm Sh}_{c} \simeq 5 \, {\rm Re}_p^{-1/2}$ when ${\rm Re}_p \lesssim 2$. A few experiments were also done for which the sphere is moving away from a wall, which reveals that the resuspension of the dust layer is much smaller because the flow it induces on the wall decreases rapidly with time and distance from the sphere, and the advective timescale associated with the flow unsteadiness is much shorter than the response time of the dust particles. However, the sphere moving away from a wall transports a large volume of fluid away from the wall, and consequently, the flow around the saltating particles leaving the ground plays a significant role in increasing the vertical mass flux of dust \citep{Eames2000}.

The flow situation of a disk approaching a dusty wall has also been considered by \cite{Khalifa2007} and \cite{Kubota2009} as a simplified model configuration for resuspension due to human walking motion. \cite{Khalifa2007} have developed an analytical model of the flow, incorporating particle detachment and transport models, to compute the trajectories of particles emanating from the floor under the descending disk. The model shows that a high velocity flow is ejected from the gap perimeter and spreads radially outward as a decelerating wall jet. This jet plays a major role in particle detachment and levitation outside the gap. A strong vortex with a significant upward motion develops outside the gap. The vortical structure and upward flow deflection are expected to be major contributors to the levitation of particles that have already detached from the disk gap. \cite{Kubota2009} have performed particle flow visualization and particle image velocimetry (PIV) measurements on a simplified model of human walking motion with a disk moving normal to the floor. On both the upstep and the downstep, particle resuspension occurs due to a high-velocity wall jet that forms between the wall and the disk, in accordance with the mechanism of particle resuspension proposed by \cite{Khalifa2007}. Large-scale ring vortex structures were formed on both the downstep and the upstep, and did not cause particle resuspension, but were highly effective at quickly moving the already resuspended particles away from the wall. By varying the seeding of the particles, it was determined that only particles underneath and toward the outer edge of the disk are resuspended. Flow visualization of particles initially seeded on the ground indicates that they are resuspended by both downward and upward motions associated with walking.

Dust resuspension was also investigated by \cite{Sajadi2013} and \cite{Elhadidi2013} for a free-falling disk under gravity onto a dusty wall. \cite{Sajadi2013} have shown that an axisymmetric vortex is generated on the disk tip as the disk falls and sheds after impacting the floor. While the effect of this ring vortex on the particles' detachment from the floor is small, it has a considerable influence on the dispersion of resuspended particles. The simulation indicated that particles are mainly resuspended from an annular area beneath the disk tip where the generated wall shear is sufficiently high.

The vortex generated in the wake of a disk of diameter $D$ with sinusoidal motion from start to stop over stroke length $L$ and time $\tau$ has been characterized by \cite{Steiner2023}. Both the ``starting'' vortex ring and the counter-rotating stopping vortex ring generated at the edge of the disk have been characterized. In the unbounded case where the disk is far from any boundaries, the maximal circulation $\Gamma_m$ and the radius $a_m$ of the starting vortex ring have been shown by \cite{Steiner2023} to follow the scaling laws $\Gamma_m \simeq 2.1 \, L^{4/3}D^{2/3} \tau ^{-1}$  and $a_m \simeq 0.1 \, L^{2/3}D^{1/3}$. These scaling laws were found to be in agreement with the analytical predictions made by \cite{Wedemeyer1961} for a starting plate with a constant velocity, but with slightly different numerical prefactors due to both different geometry and different motion law.

\cite{Steiner2025} have also performed PIV measurements and numerical simulations for the sinusoidal translation of a disk towards or away from a solid wall, down or from the minimal distance $b$. This study has shown that a disk approaching a solid bottom wall drives an outward squeezing flow in the disk-wall gap that is maximum at the edge of the disk with the gap-averaged velocity $U_{m}=V_{m} (D/4b)(1+L/b)^{-1/2}$, where $V_{m}$ is the maximum disk velocity during the translation of the disk, and that occurs during the second decelerating part of the disk motion at the dimensionless time $t^\ast = t/\tau = 1-(1/\pi){\rm acos}(1+2b/L)^{-1} > 1/2$. In the opposite case, where the disk moves away from the solid wall, the inward suction flow in the disk-wall gap is maximum with the same velocity value. Still, it occurs during the first accelerating part of the disk motion at the dimensionless time $t^\ast = (1/\pi){\rm acos}(1+2b/L)^{-1} < 1/2$. In both cases, the vortices are reinforced by the wall with the following modified scalings from the unbounded case: $|\Gamma_m| \simeq 1.3 \, L^{1.23}D^{1.05}b^{-0.28}\tau^{-1}$ and $a_m \simeq 0.068 \, L^{0.55}D^{0.59}b^{-0.14}$ for the approaching case, and $|\Gamma_m| = 1.1 \, L^{1.15}D^{1.12}b^{-0.28}\tau^{-1}$ with unchanged $a_m$ for the opposite case. Owing to the contrasting flow fields—and hence wall stresses—induced by a disk moving toward or away from a wall shown by \cite{Steiner2025}, we anticipate corresponding differences in both the onset and the spatial distribution of erosion.

 %For design or mitigation, one needs quantitative thresholds that link grain properties to the kinematics of the impulsive forcing \cite[][]{Zhang2022}.

In the present study, we aim to measure the erosion threshold of a granular bed subjected to a single linear stroke by a translating disk, in both directions towards and away from the bed.  Varying the disk diameter, stroke length, and minimum gap, we determine the critical velocity at which the first grains get eroded. The paper is organized as follows.  Section~\ref{sec:methods} details the experimental apparatus and control parameters. Section~\ref{sec:towards} considers the erosion produced by a disk approaching the bed, while Section~\ref{sec:away} characterizes the departure case.  Section~\ref{sec:conclusion} summarizes the main findings and outlines remaining questions.

%%%%%%%%%%%%%%%%%%%%%%%%%%%%%%%%%%%%%%%%%%%%%%%%%%%%%%%%%%%%%%
%%%%%%%%%%%%%%%%% EXPERIMENTAL METHODS %%%%%%%%%%%%%%%%%%%%%%%
%%%%%%%%%%%%%%%%%%%%%%%%%%%%%%%%%%%%%%%%%%%%%%%%%%%%%%%%%%%%%%

\section{Experimental methods}\label{sec:methods}

\subsection{Experimental setup}

\begin{figure}
  \centering
  \includegraphics[width=0.45\linewidth]{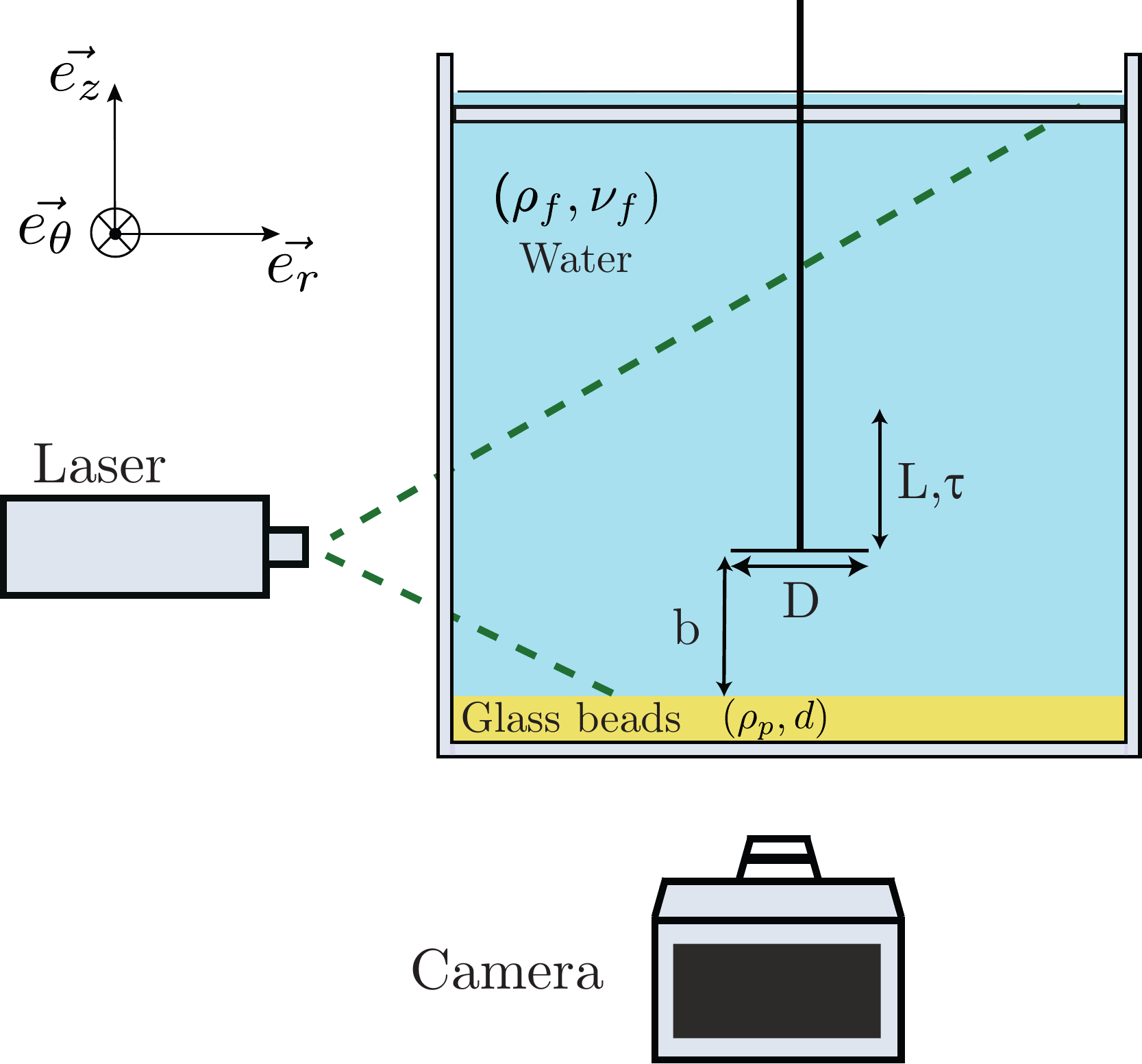}
  \caption{Schematic of the experimental setup: A rigid disk of diameter $D$ translates vertically over a stroke $L$ in time $\tau$, either descending to the minimum gap $b$ or ascending from it.} %
  \label{fig:Figure_1} 
\end{figure}

A schematic of the experimental setup is shown in Fig.~\ref{fig:Figure_1}. This setup has been used in previous studies of fluid flows around a translating disk \cite[]{Steiner2023,Steiner2025}. In summary, the experiments were performed in a square glass tank (side $40\,{\rm cm}$ and height $60\,{\rm cm}$) with the water level set to $40\;\mathrm{cm}$. A transparent acrylic lid was placed a few millimeters beneath the free surface to suppress surface waves. At the bottom, a $1\,{\rm cm}$ thick, horizontally leveled bed of glass beads (mean diameter $d\simeq 250 \pm 50\,\mu{\rm m}$, density $\rho_{p}=2500\,{\rm kg\,m^{-3}}$) was prepared. Before each erosion experiment, the surface was smoothed with a straight edge to ensure a repeatable packing and a flat surface.

A circular rigid PVC disk with a diameter $D \in[5,\, 15]\,{\rm cm}$, small enough compared to the tank dimensions so that lateral wall effects remain negligible, and with a small thickness $e=2\,{\rm mm}$ ($e/D \lesssim 0.05$) was attached to a vertical shaft (1 cm diameter) driven by an AC servomotor (ECMA-C20807RS, Delta Electronics). An eccentric cam converts the continuous rotation of the motor into a sinusoidal translation of the vertical shaft. The stroke length $L$ has been varied here in the range $2\,{\rm cm} \leq L \leq 5.2\,{\rm cm}$ and the minimum distance $b$ between the disk and the granular bed in the range $0.2\,{\rm cm} \leq b \leq 2\,{\rm cm}$. 

Throughout this study, the motion of the disk is a single translation over the time $\tau$, during which the servomotor performs half a revolution at constant angular velocity $\pi/\tau$, so that the disk accelerates and decelerates sinusoidally. The disk therefore travels the prescribed distance $L$ in a time $\tau$ and then comes to rest. 
Introducing the dimensionless time $t^{\ast}=t/\tau$, the prescribed vertical position $h$ and velocity $V$ of the disk are  
\begin{equation}
  h(t^{\ast})=b+\frac{L}{2}\!\left[1\mp\cos(\pi t^{\ast})\right],\qquad
  V(t^{\ast})=\pm V_{m}\sin(\pi t^{\ast}),
  \qquad 0\le t^{\ast}\le1,
  \label{eq:disk-motion}
\end{equation}
with $V_{m}=\pi L /(2\tau)$ the maximum translational velocity at mid-stroke $L/2$ and mid-time $t^{\ast}=1/2$. The upper (resp. lower) sign applies when the disk starts from the bottom (resp. top) position.

\begin{table}
\centering
\setlength{\tabcolsep}{12pt}
\begin{tabular}{ c c c c c } 
%\multicolumn{7}{c}{}\\
 \hline \hline
 $L$ (cm) & $D$ (cm) & $L/D$ & $b/D$ &Symbols\\
 \hline \hline
 2 & 10 & 0.2 & 0.02-0.2 & \color{cyan}{$\medbullet$} \\
 2.8 & 10 & 0.28 & 0.02-0.2 & \color{blue}{$\medbullet$} \\
 3.6 & 10 &  0.36 & 0.02-0.2 & \color{green}{$\medbullet$} \\
 4.4 & 10 &  0.44 & 0.02-0.2 & \color{yellow}{$\medbullet$} \\
 5.2 & 10 &  0.52 & 0.02-0.2 & \color{orange}{$\medbullet$} \\
 2.8 & 5 & 0.56 & 0.04-0.4 & \color{blue}{$\blacklozenge$} \\
 2.8 & 7.5 & 0.37 & 0.027-0.27 & \color{blue}{$\blacktriangle$} \\
 2.8 & 12.5 & 0.22 & 0.016-0.16 & \color{blue}{$\blacktriangledown$} \\
 2.8 & 15 & 0.19 & 0.013-0.13 & \color{blue}{$\blacksquare$} \\
 \hline \hline
\end{tabular}
 \caption{Ranges of the parameters $L$ and $D$ explored in the different sets of experiments together with the corresponding symbols used in the figures.}
 \label{tab:ParametersThreshold}
\end{table}

The parameter ranges explored across all experiments are summarized in Table~\ref{tab:ParametersThreshold}, along with the corresponding data symbols used in the figures. The first five sets of experiments correspond to the same disk diameter, $D = 10$ cm, with different stroke lengths, $L$, and the data will be represented by different-colored circles. The last four sets of experiments, together with the second set, correspond to different $D$ values for the same $L = 2.8$ cm, and the data will appear as blue symbols of different shapes.

\subsection{Determination of the erosion threshold}\label{sec:threshold}

Two complementary procedures were used to identify the onset of erosion of the granular bed. First, direct visualization was employed: for a set of control parameters $(L,\,D,\,b)$, the disk velocity $V$ was increased by decreasing the translation time $\tau$. The threshold of erosion corresponds to the critical value $\tau_c$ at which the very first grains were observed to be eroded from the bed. The corresponding maximum velocity $V_{m}^{\,c}=\pi L/(2\tau_{c})$ is taken as the erosion threshold. Each erosion threshold was measured three times in independent experiments, and the resulting standard deviation of the critical velocity was typically about 5–10\%.

A second method used relied on laser profilometry: The threshold was defined as the smallest $V_{m}$ for which the mean depth change exceeded $10\,\%$ of the initial layer thickness. Although the absolute values obtained by the two methods differed slightly, they are consistent for comparative purposes. Erosion thresholds generated by the radial flow before the disk stops—during both approach and retreat—were determined by direct visualization in sections \ref{sec:Eros_squeez} and \ref{sec:DPB}, whereas the laser-profilometry criterion was applied exclusively in section \ref{sec:stop} to the vortex-driven erosion that develops after the disk has stopped, whose strongly unsteady and fluctuating nature precludes a reliable threshold from instantaneous images.

\subsection{Control parameters and non‑dimensional groups}\label{sec:dim}

The problem is fully characterized by the geometric and kinematic parameters $(D,L,b,V_{m})$ of the disk motion, the density $\rho$ and dynamic viscosity $\eta$ of the fluid, and the grain parameters $(\rho_{p},d)$ together with the gravitational acceleration $g$. Thus, the six following independent dimensionless parameters can be considered:
\begin{equation}
  \text{Re}=\frac{\rho \, V_{m}\,D}{\eta},\quad
  \frac{L}{D},\quad
  \frac{b}{D},\quad
  \frac{\Delta\rho}{\rho},\quad
  \text{Ar}=\dfrac{\rho\,\Delta\rho\,g\,d^{3}}{\eta^{2}},\quad
  \text{Sh}=\dfrac{\rho V_{m}^{2}}{\Delta\rho\,g\,d},
  \label{eq:dimensionless}
\end{equation}
where $\Delta \rho = \rho_p - \rho$ is the grain-fluid density difference. In the present study, the density ratio is fixed at $\Delta \rho / \rho \simeq 1.5$. The Archimedes number $\rm{Ar} \simeq 230$ indicates that grain settling occurs in the inertial regime. The remaining dimensionless groups cover the following ranges: the Reynolds number Re spans $10^3 \lesssim \rm{Re} \lesssim 10^{5}$, the relative stroke of the disk lies within $0.19 \lesssim L/D \lesssim 0.56$, the normalized gap $b/D$ ranges from 0.013 to 0.4, and the Shields number Sh varies approximately between $10^{-2} \lesssim \rm{Sh} \lesssim 10^2$. The range of Re ensures that the disk‑generated vortical flow is mostly in the inertial regime, while the broad variation of $b/D$ together with the variation of $L/D$ allows us to investigate the respective roles of the different flow features on the erosion threshold.

%%%%%%%%%%%%%%%%%%%%%%%%%%%%%%%%%%%%%%%%%%%%%%%%%%%%%%%%%%%%%%
%%%%%%%%%% EROSION - MOTION TOWARDS THE BED %%%%%%%%%%%%%%%%%%
%%%%%%%%%%%%%%%%%%%%%%%%%%%%%%%%%%%%%%%%%%%%%%%%%%%%%%%%%%%%%%

\begin{figure}
  \centerline{\includegraphics[width=1\linewidth]{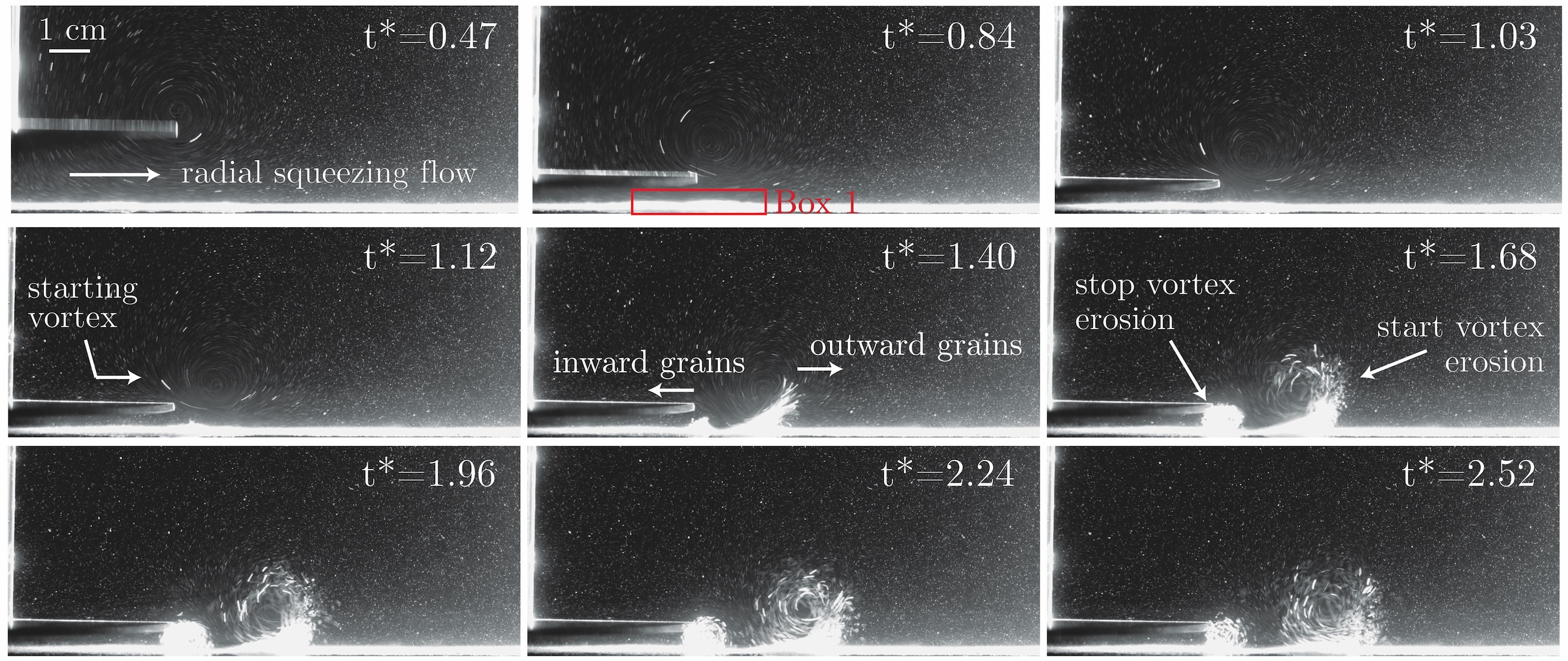}} 
  \caption{Pictures taken at nine successive dimensionless times $t^{\ast}=t/\tau$ illustrating the erosion of a granular bed induced by the flow generated when a disk of diameter $D = 10\,{\rm cm}$ translates along the stroke length $L = 2.8\,{\rm cm}$ during the travel time $\tau \simeq 0.21\,{\rm s}$ down to the minimum/final gap distance $b = 0.5\,{\rm cm}$ from the granular bed ($V_m = 21$ cm/s  $\simeq 2V_m^c$.)}
  \label{fig:Figure_2}
\end{figure}

\section{Erosion by a disk moving towards a granular bed}
\label{sec:towards}
%%%%%%%%%% Phenomenology %%%%%%%%%%%%%%%%%%
\subsection{Phenomenology}\label{sec:phenomenology}

Figure~\ref{fig:Figure_2} presents a time-lapse sequence from an experiment performed well above the erosion threshold, illustrating the two distinct mechanisms by which erosion arises when the disk moves downward.

\begin{figure} 
  \centerline{\includegraphics[width=1.1\linewidth]{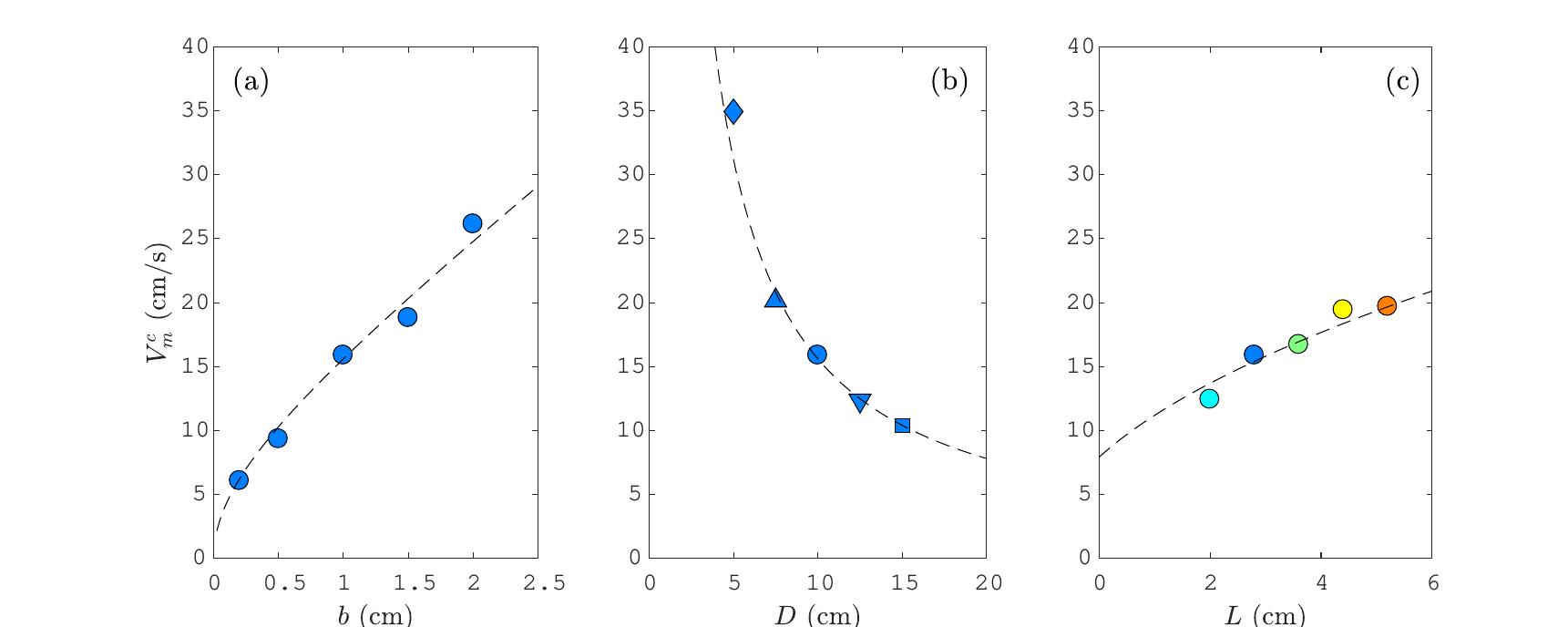}}
  \caption{Critical velocity $V_m^{c}$ of the disk at the onset of erosion measured before the disk stops ($t^{*}<1$) for a disk translating towards the granular bed as a function of (a) the minimal/final distance from the granular bed $b$ for $L = 2.8\,{\rm cm}$ and $D =10\,{\rm cm}$, (b) the disk diameter $D$ for $L = 2.8\,{\rm cm}$ and $b = 1\,{\rm cm}$, and (c) the stroke length $L$ for $D = 10\,{\rm cm}$ and $b = 1\,{\rm cm}$. Dashed lines are the best fits from Eq.~\eqref{eq:Vmcrit} corresponding to (a) $V_m^c = 8 b \sqrt{1+2.8/b}$, (b) $V_m^c = 156 / D$, and (c) $V_m^c = 7.9 \sqrt{1 + L}$ with (a) $U_m^c = 19.75$ cm/s, (b) $U_m^c = 20.0$ cm/s, and (c) $U_m^c = 19.75$ cm/s.}
  \label{fig:Figure_3}
\end{figure}

The first mechanism occurs when erosion takes place while the disk is still moving ($t^{\ast}<1$). As the disk approaches the bed, the fluid trapped beneath is expelled radially. The resulting outward radial flow, confined in the thin gap, shears the grains near the edge of the disk where the velocity is the largest \citep{Steiner2025}. Here, between $t^{\ast} \simeq 0.4$ and $t^{\ast} \simeq 1$, a small ridge forms at $r\simeq D/2$ (see box 1 in figure~\ref{fig:Figure_2} at $t^*=0.84$) and drifts outward, revealing that only a narrow annulus of grains is eroded while the disk is in motion. At the same instant, the starting vortex is already visible in the wake of the disk, but its core is sufficiently far from the bed for the outward flow beneath the disk to remain predominantly radial and axisymmetric. Therefore, the first erosion threshold is governed by the wall shear generated by this squeezing flow in the gap.

The second mechanism is through vortex-driven erosion after the disk stops ($t^{\ast}>1$). Indeed, once the stroke ends, the starting vortex detaches, flows around the disk, and impinges on the bed. The interaction of this vortex with the edge of the disk detaches a counter-rotating stopping vortex. Together, they form a dipole that translates parallel to the surface of the granular bed. In the present example, from $t^{\ast}=1.12$ to $1.68$, grains are entrained simultaneously outward by the starting vortex and inward by the stopping vortex, producing the bidirectional transport highlighted in panels $t^{\ast}=1.40$ and $1.68$. The amount of suspended granular material grows until the vortices weaken, and by $t^{\ast}=2.52$, they can no longer sustain the grains in suspension, which then resettle onto the bed. Because the circumferential velocities in the starting and stopping vortex after the disk stops are comparable and larger than the outward flow velocity before the disk stops, the erosion mechanism by the impact of a vortex ring dipole typically imposes a higher wall stress than the squeezing radial flow. It therefore controls a separate and smaller erosion threshold.

In summary, when the disk moves towards the bed, the transient erosion arises first from the shear induced by the outward squeezing flow in the gap and then, if the disk stops sufficiently close to the bed, from the following impact of the starting–stopping vortex pair. In the following, we quantify the different erosion thresholds and describe how they scale with the governing parameters.

%%%%%%%%%% EROSION - Radial squeezing flow %%%%%%%%%%%%%%%%%%

\subsection{Erosion by the radial squeezing flow before the disk stops} \label{sec:Eros_squeez}

When the disk moves toward the granular bed, the fluid confined in the narrow gap is expelled radially. The associated shear can erode the granular bed before the disk comes to a stop. We quantify here the critical disk velocity $V_{m}^{c}$ corresponding to the first onset of erosion and demonstrate how it depends on the control parameters.

Figures~\ref{fig:Figure_3}(a)-(c) report the values of $V_{m}^{c}$ obtained by progressively increasing the imposed stroke velocity until a few grains begin to be eroded and transported radially outwards when the parameters $b$, $D$, and $L$ are varied separately. The critical velocity $V_{m}^{c}$ is found to increase with the final gap $b$ [Fig.~\ref{fig:Figure_3}(a)], decreases with the disk diameter $D$ [Fig.~\ref{fig:Figure_3}(b)], and increases with the stroke length $L$ [Fig.~\ref{fig:Figure_3}(c)]. These trends directly reflect the way the radial flow required to conserve mass in the gap varies with the three parameters. To develop a scaling law using the maximum radial velocity, \cite{Steiner2025} assumed that the gap thickness $h(t)$ remains small compared with $D$. Then, the incompressibility condition leads to a gap-averaged radial velocity profile $U$ that is maximum at $r=D/2$. The maximum value over time is $U_{m}=V_{m} \left (D/4b \right ) \left (1+L/b \right )^{-1/2}$, where $V_{m}$ is the maximum disk velocity during the translation of the disk. The shear at the granular bed required for erosion is directly related to the radial flow and thus to $U_{m}$. Assuming that the gap-averaged radial velocity $U_{m}^{c}$ should remain constant for a given size and density of grains, the critical disk velocity $V_{m}^{c}$ can be written as
\begin{equation}
V_{m}^{c}=U_{m}^{c}\,\frac{4b}{D}\,\sqrt{1+\frac{L}{b}}.
\label{eq:Vmcrit}
\end{equation}
Dashed lines in Fig.~\ref{fig:Figure_3} show the \(V_m^{c}(b,D,L)\) best fits of the data from Eq.~\eqref{eq:Vmcrit} that give very close fitting values for $U_m^{c}$. The very good fit of the data for the corresponding set of parameters, shown in Fig.~\ref{fig:Figure_3}, suggests that the measured erosion thresholds should obey this relationship.

\begin{figure}
  \centerline{\includegraphics[width=\linewidth]{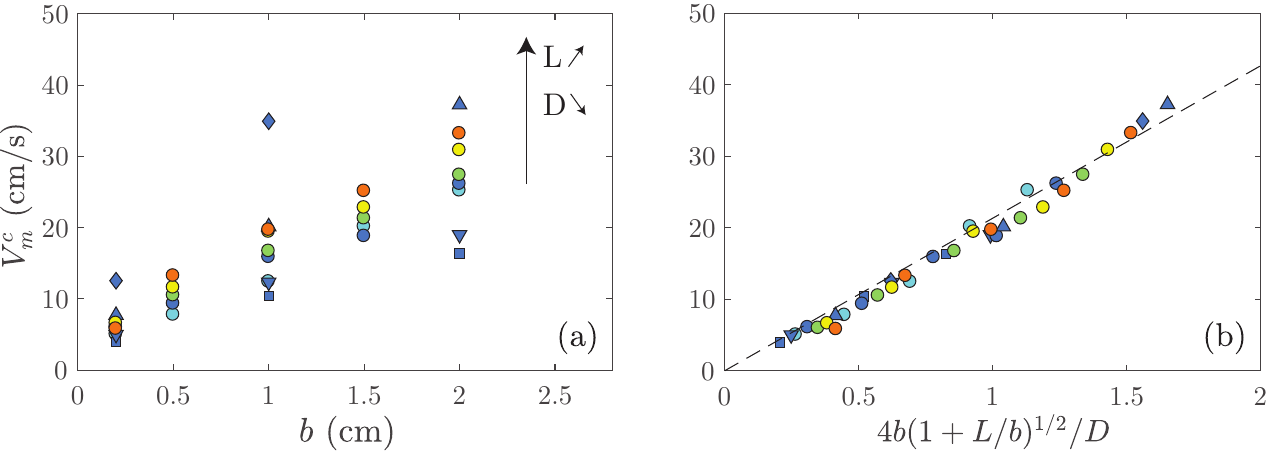}}
  \caption{Critical velocity of the disk at the onset of erosion by the radial outflow before the disk stops ($t^*<1$), $V^{c}_{m}$, as a function of (a) the minimum distance $b$ and (b) the parameter $4b(1+L/b)^{1/2}/D$. The dashed line corresponds to the best fit of the data by Eq. (3.1) with $U_{m}^{c}=21.3\,{\rm cm\,{s}^{-1}}$.}
  \label{fig:Figure_4}
\end{figure}

 The critical disk velocity $V_m^{c}$ required to dislodge the very first grains for all stroke lengths $L$, disk diameters $D$, and final gaps $b$ explored here with the color code given in Table~\ref{tab:ParametersThreshold} is shown in Fig.~\ref{fig:Figure_4}(a). The observed global trend is again that $V_m^{c}$ grows with $b$, showing that a wider gap weakens the outward radial flow and requires a faster stroke for the shear at the granular bed to reach the erosion threshold. To account for the combined influence of \(L\), \(D\), and \(b\) on the measured thresholds, the critical disk velocity \(V_m^{c}\) is now plotted as a function of \(4b\sqrt{1+L/b}/D\) in Fig.~\ref{fig:Figure_4}(b). All data collapse remarkably well onto a linear master curve as predicted from Eq.~\eqref{eq:Vmcrit} with the fitted value $U_m^{\,c}=(21.3 \pm 2.2) \, \mathrm{cm\,s^{-1}}$. Thus, Fig.~\ref{fig:Figure_4}(b) demonstrates that the first erosion mode—driven solely by the outward squeezing flow before the disk stops—is fully captured by this gap-averaged kinematic model.

\begin{figure}
  \centerline{\includegraphics[width=1.0\linewidth]{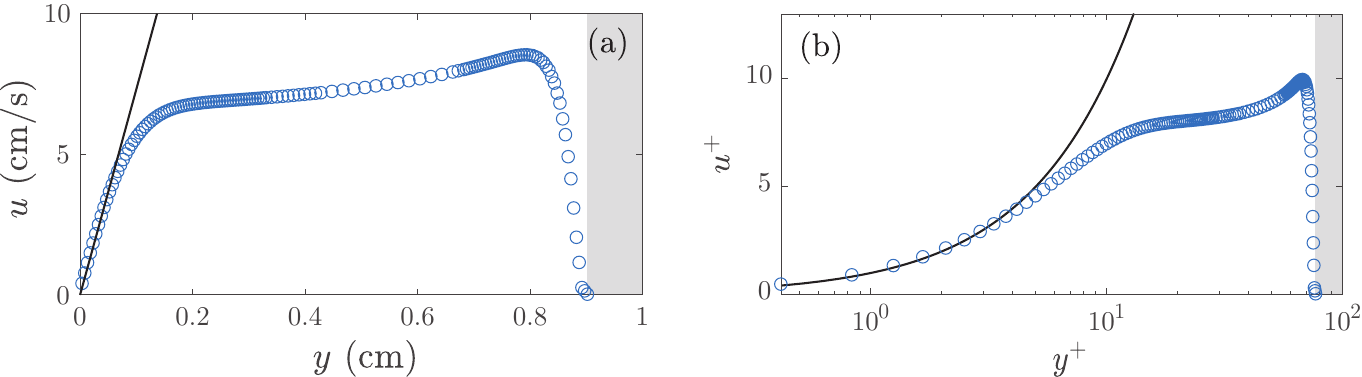}}
  \caption{(a) Velocity profile $u(y)$ across the gap between the disk ($y = h$) and the wall ($y = 0$), sampled at the disk edge ($r=D/2$) at the dimensionless time $t^{\ast}=0.76$ where and when the fluid velocity $U_m^c$ is maximum, for $L = 2.8\,\text{cm}$, $D = 10\,\text{cm}$, $b = 0.5\,\text{cm}$, and $\tau = 1.25\,\text{s}$ (Re $\simeq 3.5 \times 10^3$). The data ($\textcolor{blue}{\circ}$) are obtained from the numerical simulations of \cite{Steiner2025}. The shaded region corresponds to the location of the disk. The solid line is  $u = 73.1 y$ corresponding to $u_{\tau} = 8.5$ mm/s and $\delta_{v}= 1.2 \times 10^{-4}$ m. (b) Corresponding dimensionless velocity profile in wall units $u^+ = u/u_{\tau}$ as a function of $y^+ = y/\delta_{v}$. The solid line is $u^+ = y^+$.}
  \label{fig:Figure_5}
\end{figure}

Velocity fluctuations and the organization of the viscous sublayer are known to control the onset of particle motion \citep{Quibeuf2020}. To estimate the velocity at the grain scale, we use the direct numerical simulations (DNS) of \cite{Steiner2025}, conducted near a rigid wall. Assuming that grains at incipient motion do not perturb the near-wall velocity, we relate the gap-averaged velocity \(U_m\) to a representative wall-scale velocity.  Figure~\ref{fig:Figure_5}(a) displays the vertical profile of the radial velocity \(u(y)\) across the gap $h$ at the disk rim \(r=D/2\) and dimensionless time \(t^{\ast}=0.76\) when the gap-averaged velocity \(U_m\) reaches its maximum. The resulting profile exhibits the canonical features of turbulent plane-channel flow: a nearly uniform velocity at the center and steep velocity gradients confined to thin boundary layers adjacent to the disk and the wall.

To obtain the wall-friction velocity, we proceed as follows. First, the instantaneous wall shear stress $\tau_{w}$ is extracted from the DNS by evaluating $\tau_{w}=\eta (\partial u/\partial y)_{y=0}$, then we deduce the friction velocity $u_{\tau}=(\tau_{w}/\rho)^{1/2}$, which represents the characteristic velocity at the viscous sublayer scale. For the case shown in Fig.~\ref{fig:Figure_5}, the friction velocity is $u_{\tau}=8.5~\mathrm{mm/s}$. The thickness of the viscous sublayer is estimated as $\delta_{v}=\nu/u_{\tau}$, which in this case gives $\delta_{v}=1.2\times 10^{-4}~\mathrm{m}$. The dimensionless velocity \(u^+ = u/u_{\tau}\) is plotted in wall units as a function of \(y^+ = y/\delta_{v}\) in Fig.~\ref{fig:Figure_5}(b). As expected, the velocity increases linearly from the wall within the viscous sublayer when $y^+ \lesssim 10$, which confirms the validity of the extracted friction velocity $u_{\tau}$. 

\begin{figure}
  \centerline{\includegraphics[width=\linewidth]{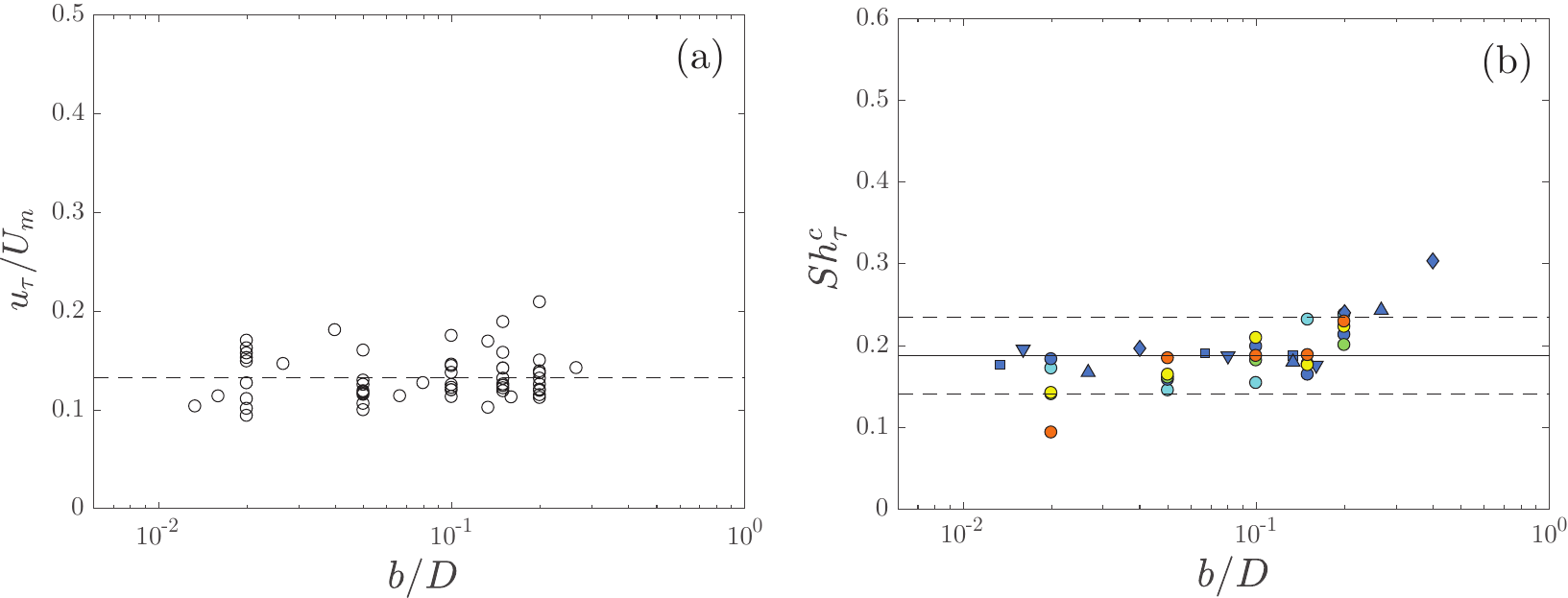}}
  \caption{(a) Velocity ratio $u_\tau/U_m$ as a function of the dimensionless gap $b/D$, obtained from the numerical simulations of \cite{Steiner2025}. The horizontal dashed line corresponds to the mean value $u_\tau/U_m \simeq 0.133$. (b) Local Shields number $\rm{Sh}^c_{\tau}$ defined from the friction velocity $u_{\tau}$ at the erosion threshold plotted as a function of $b/D$. The solid line marks the mean value $\rm{Sh}_{\tau}^{c}\simeq 0.188$, while the dashed lines indicate a $\pm 25\%$ deviation from this mean.}
  \label{fig:Figure_6}
\end{figure}

The friction velocity \(u_{\tau}\) is compared to the vertically averaged gap velocity \(U_{m}\) for all relative disk–wall gaps $b/D$ investigated in Fig.~\ref{fig:Figure_6}(a). The ratio \(u_{\tau}/U_{m}\) remains essentially constant -- $0.13 \pm 0.04$ -- over the entire data set, indicating that the friction velocity \(u_{\tau}\) is proportional to $U_{m}$ which justifies using a single conversion factor between the two when predicting erosion thresholds. The particle Reynolds number, $Re_{p}=u_{\tau} d / \nu$ is of the order of $\simeq 7$ at the onset of erosion over all conditions tested. This value is significantly greater than unity, indicating that viscous effects in the near-wall flow are negligible compared to inertia at the grain scale. Consequently, we define here a local Shields number $\rm{Sh}_{\tau}$ based on the friction velocity $u_{\tau}$ from the radial squeezing flow,
\begin{equation}
\text{Sh}_{\tau}=\frac{\rho {u_\tau}^{2}}{\Delta\rho\, g\, d}.
\end{equation}
The critical values $\rm{Sh}_{\tau}^c$ for the local Shields number are shown in Figure~\ref{fig:Figure_6}(b) as a function of the gap/disk size ratio $b/D$, which shows that they remain nearly constant around $0.19\pm25\%$ over more than one decade of $b/D$. This value is relatively close to the value $\rm{Sh}_{\tau}^c = 0.13 \pm 0.01$ known for steady homogeneous flows \citep{Loiseleux2005,Ouriemi2007}.

In summary, the erosion driven by the radial squeezing flow is set by the critical gap-averaged radial velocity $U_{m}^{c}\simeq 21\,$cm\,s$^{-1}$ corresponding to the critical friction velocity $u_\tau^{c} \simeq 2.8\,$cm\,s$^{-1}$ for the present glass beads, which leads to the critical values $\rm{Sh}_{\mathrm{R}}^{c} \simeq 11 $ and $\rm{Sh}_{\tau}^c \simeq 0.19$ for the inertial Shields number based on $U_{m}^{c}$ and $u_\tau^{c}$, respectively. These results provide a reliable predictive criterion for design situations where rapid gap compression mobilises sediment.

%%%%%%%%%% EROSION - After Stops %%%%%%%%%%%%%%%%%%

\subsection{Erosion by vortex ring dipole after the disk stops} \label{sec:stop}

\begin{figure}
  \centerline{\includegraphics[width=\linewidth]{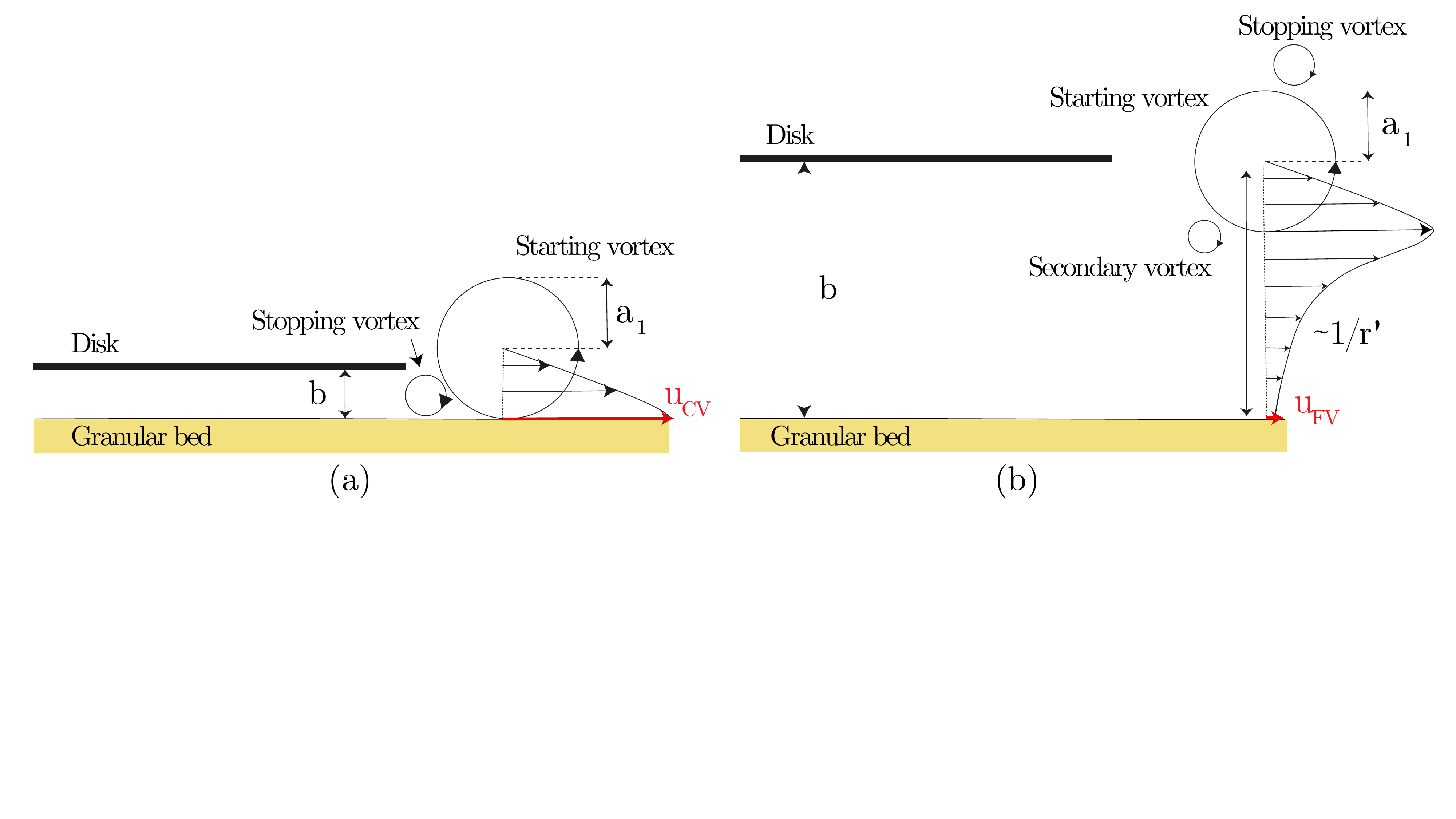}}
  \vspace{-2.8cm}\caption{Schematic of (a) the impact of the dipole formed by the starting vortex and the stopping vortex on the granular bed when the disk stops close to the bed ($b/a_1 \lesssim 1$) and of (b) the near-wall velocity behavior when the disk stops farther from the bed ($b/a_1\gtrsim 1$).}
  \label{fig:Figure_7}
\end{figure}

Once the disk has completed its stroke ($t^{\ast}>1$), the starting vortex ring, generated in its wake during the translation, drifts downward and interacts with the rim of the disk. This interaction detaches a counter-rotating stopping vortex which, together with the starting vortex, forms a dipole of vortex rings that propagates parallel to the granular bed surface as illustrated on the schematics of Fig~\ref{fig:Figure_7}. When the disk stops sufficiently close to the bed ($b/a_1\lesssim 1$, where $a_1 = a(t^*=1)$ is the core radius of the starting vortex at the disk stop) as illustrated in Fig~\ref{fig:Figure_7}(a), the dipole impinges on the grains almost immediately, entraining sediment both outward for the starting vortex and inward for the stopping vortex, thus generating bidirectional transport described in Fig.~\ref{fig:Figure_2}.  

When the disk stops far from the bed ($ b/a_1 \gtrsim 1$), the stopping vortex flows around the starting vortex without impacting the bed directly. Therefore, erosion is delayed and results either from the slowly decaying velocity field of the starting vortex itself or from the later impact of secondary stopping vortices [Fig.~\ref{fig:Figure_7}(b)].  
These two scenarios will be referred to as CV for the close vortex case ($b/a_1\lesssim 1$) and FV for the far vortex case ($b/a_1\gtrsim 1$), respectively. Thus, two distinct situations, depending on the ratio $b/a_1$, will be considered when analyzing the erosion thresholds after the disk comes to rest.

\begin{figure}
\centerline{\includegraphics[width=0.95\linewidth]{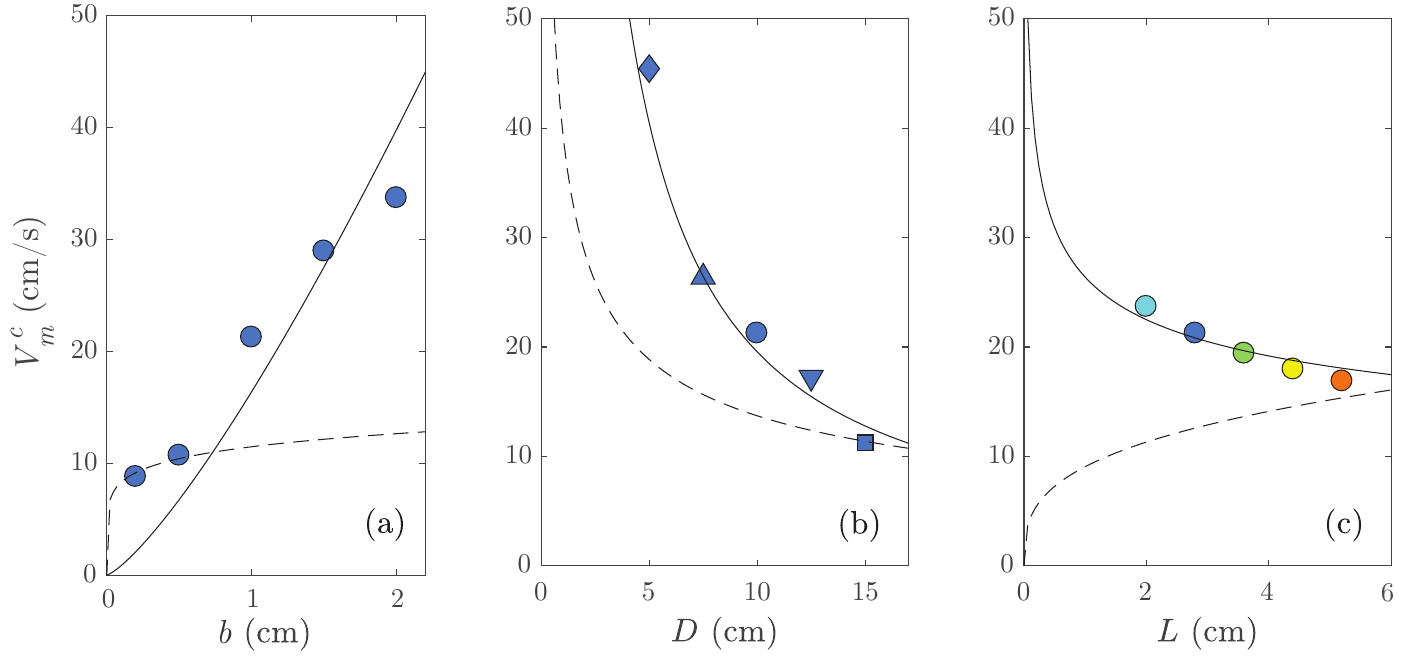}}
\caption{Critical velocity of the disk at the onset of erosion occurring once the disk has come to rest ($t^{*}>1$), $V_m^{c}$, measured for a disk released from its upper position as a function of (a) the final distance to the granular bed, $b$ ($L = 2.8\,{\rm cm}$, $D = 10\ {\rm cm}$), (b) the disk diameter, $D$ ($L = 2.8\ {\rm cm}$, $b = 1\ {\rm cm}$), and (c) the stroke length, $L$ ($D = 10\ {\rm cm}$, $b = 1\ {\rm cm}$). Dashed (CV) and solid lines (FV) are the best fits of the data from Eq. (3.6) with similar $u_{\text{CV}}^c$ or $u_{\text{FV}}^c$ values, respectively, corresponding to (a) $V_m^c \simeq 0.28 u_{\text{CV}}^c \, b^{0.14}$ with $u_{\text{CV}}^c \simeq 41$ cm/s and $V_m^c \simeq 0.63 u_{\text{FV}}^c \, b^{1.28}$ with $u_{\text{FV}}^c \simeq 26$ cm/s, (b) $V_m^c \simeq 0.81 \, u_{\text{CV}}^c/D^{0.46}$ with $u_{\text{CV}}^c \simeq 49$ cm/s and $V_m^c \simeq 7.1 u_{\text{FV}}^c/D^{1.05}$ with $u_{\text{FV}}^c \simeq 31$ cm/s and (c) $V_m^c \simeq 0.20 \, u_{\text{CV}}^c L^{0.32}$ with $u_{\text{CV}}^c \simeq 45$ cm/s and $V_m^c \simeq 0.80 \, u_{\text{FV}}^c/L^{0.23}$ with $u_{\text{FV}}^c \simeq 33$ cm/s. }
  \label{fig:Figure_8}
\end{figure}

The critical disk velocity $V_{m}^{c}$ required to trigger vortex-driven erosion after the disk stops is reported in Fig.~\ref{fig:Figure_8}(a)-(c). For a fixed disk diameter $D$ and stroke length $L$, $V_{m}^{c}$ increases monotonically with the minimum distance $b$ to the bed [Fig.~\ref{fig:Figure_8}(a)], meaning that a faster stroke is needed when the vortices must travel farther before hitting the grains. In contrast, $V_{m}^{c}$ decreases with the disk diameter $D$ [Fig.~\ref{fig:Figure_8}(b)] because the circulation of the starting vortex scales roughly with $D$ \citep{Steiner2025}, so that a larger disk generates a stronger vortex for the same disk velocity. Finally, $V_{m}^{c}$ is found to decrease with the stroke length $L$ [Fig.~\ref{fig:Figure_8}(c)], which is opposite to what we observed for the squeezing-flow threshold discussed in the previous section. This difference can be explained by a longer acceleration phase, which allows the vortex to build up more circulation, which scales roughly with $L^{1.2}$ \citep{Steiner2025}, and thus leads to bed erosion at a lower disk velocity.

\begin{figure}
  \centerline{\includegraphics[width=\linewidth]{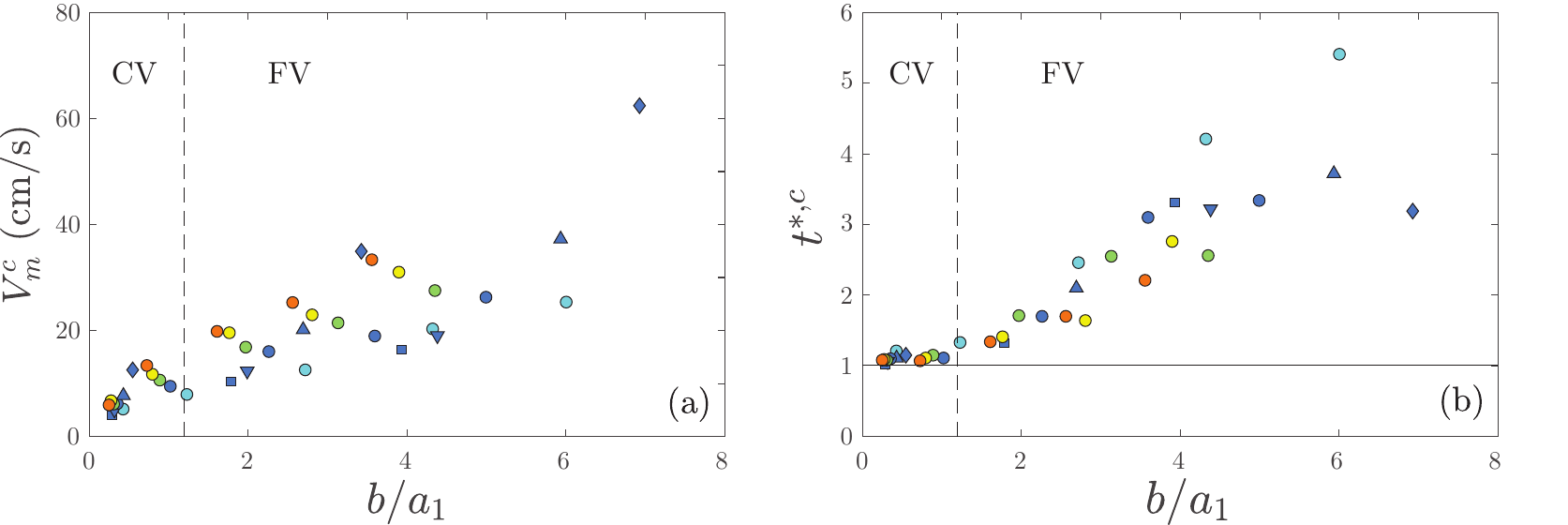}}
\caption{(a) Velocity of the disk at the erosion threshold as a function of $b/a_1$. (b) Dimensionless time $t^{*}_{c}$ at which the first grains begin to be eroded; the horizontal line marks the instant when the disk comes to rest. In both panels, the vertical dashed lines show the transition $b/a_1=1.2$ between CV and FV. }
  \label{fig:Figure_9}
\end{figure}

All the critical values $V_{m}^{c}$ for vortex-driven erosion are shown in Fig.~\ref{fig:Figure_9}(a) as a function of the size ratio $b/a_1$ of the gap with the core radius $a_1$ of the vortex at the disk stop. There is substantial dispersion, but the global trend is that $V_{m}^{c}$ increases with larger $b/a_1$. The critical dimensionless time $t^{*}_{c}$ at which the first grains begin to be eroded just above the threshold is shown in Fig.~\ref{fig:Figure_9}(b). The data show that $t^{*}_{c} \simeq 1$ for $b/a_1 \lesssim 1.2$ and that $t^{*}_{c} > 1$ for $b/a_1 \gtrsim 1.2$, which corresponds to the CV and FV erosion scenarios where the vortex erosion arises at the disk stop or later, respectively.

To estimate the forcing and corresponding velocity near the granular bed, we consider that the shear applied by the vortices on the granular bed can be estimated by the circumferential velocity of the vortex close to the granular bed.  
The two following expressions are appropriate depending on the position of the vortex core at $t^*=1$:
\begin{equation}
u_{\text{CV}}= \frac{\Gamma_1}{2\pi a_1}, 
\qquad
u_{\text{FV}}= \frac{\Gamma_1}{2\pi b},
\label{eq:Udef}
\end{equation}
for $b/a_1 \lesssim1.2$ and $\gtrsim1.2$ respectively, where $\Gamma_1 = \Gamma(t^*=1)$ and $a_1$ denote the circulation and core radius of the starting vortex at the disk stop. For $b/a_1\lesssim1.2$ (CV), the stopping vortex is trapped between the disk and the bed so that the dipole impinges directly on the bed, and the first expression in Eq.~\eqref{eq:Udef} therefore sets the relevant shear. When $b/a_1\gtrsim1.2$ (FV), the starting vortex remains at an average distance $b$ from the grains and its $1/r'$ velocity decay controls the erosion at $r' \simeq b$, leading to the second expression in Eq.~\eqref{eq:Udef}.

\begin{figure}
  \centerline{\includegraphics[width=\linewidth]{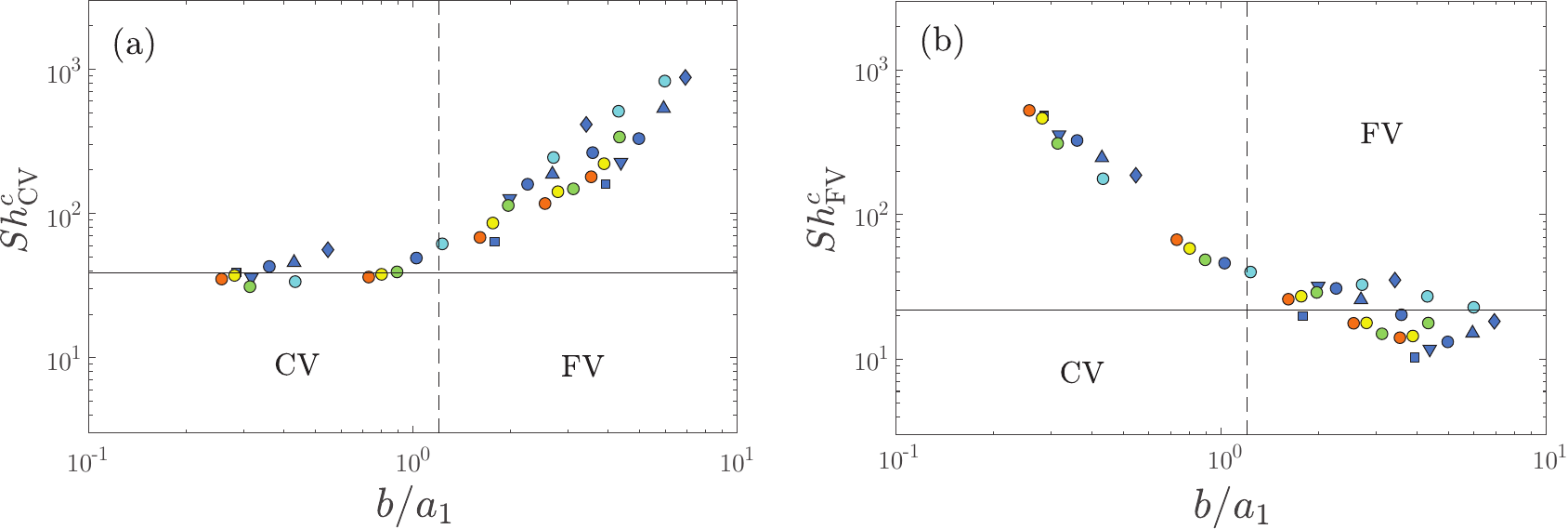}}
\caption{(a) Critical Shields number $Sh_{\mathrm{CV}}^{c}$, defined with the local velocity $u_{CV}$ induced by the impact of the starting vortex ring, as a function of $b/a_1$. (b) Critical Shields number $Sh_{\mathrm{FV}}^{c}$, defined with the local velocity $u_{FV}$ induced by the starting vortex as a function of $b/a_1$. The vertical dashed lines mark the transition at $b/a_1=1.2$, while the horizontal line indicates in (a) the mean value $\overline{\rm{Sh}_{\mathrm{CV}}^{c}} = 39$ and in (b) the mean value $\overline{\rm{Sh}_{\mathrm{FV}}^{c}} = 22$. }
  \label{fig:Figure_10}
\end{figure}

The circulation $\Gamma_1$ and radius $a_1$ of the starting vortex at the disk stop measured by \cite{Steiner2023PhD} and given in Eq.~\eqref{eq:circulation_radius_toward} in Appendix~\ref{appendix:A.2.1} are the following: $|\Gamma_1|~\simeq 1.1 \, L^{1.23}D^{1.05}b^{-0.28}\tau^{-1}$ and $a_1 \simeq 0.064 \, L^{0.55}D^{0.59}b^{-0.14}$. They follow the same scalings as the maximal circulation $\Gamma_m$ and radius $a_m$ reported in \cite{Steiner2025} with numerical prefactors only a little smaller: 1.1 instead of 1.3 for the circulation and 0.064 instead of 0.068 for the radius, corresponding to a decrease of only 20\%  and 10\%, respectively. The corresponding expressions for the velocity induced by the vortex in the vicinity of the grains, $u_{\text{CV}}$ and $u_{\text{FV}}$, are thus
\begin{equation}
u_{\text{CV}} \simeq \frac{17 \, L^{0.68} \, D^{0.46}}{2\pi \,b^{0.14} \,\tau}, 
\qquad
u_{\text{FV}} \simeq \frac{1.1 \, L^{1.23} \, D^{1.05}}{2\pi \, b^{1.28} \, \tau}.
\label{eq:Uscaling}
\end{equation}
Note that the two scalings of Eq.~\eqref{eq:Uscaling} are very different. A quantitative comparison between the experimental thresholds and the theoretical predictions is provided in Fig.~\ref{fig:Figure_8}(a)-(c). The dashed lines denote the confined-vortex (CV) regime, in which the stopping vortex is confined between the disk and the bed; the solid lines denote the far-vortex (FV) regime, in which the far-field action of the vortex erodes the granular bed from a distance of order $b$. In each case, the curves represent best fits obtained from Eq.~\eqref{eq:Uscaling}, with fitted coefficients reported in the caption of Fig.~\ref{fig:Figure_8}. The good agreement between experiments and the model based on Eq.~\eqref{eq:Uscaling} confirms the relevance of the scaling arguments for both regimes and highlights the distinct physical mechanisms controlling erosion depending on the disk–bed separation. It is worth noting that in Fig.~\ref{fig:Figure_8}(a), the first two data points correspond to the confined-vortex (CV) regime, while the remaining points fall in the far-vortex (FV) regime. By contrast, all the data shown in Figs.~\ref{fig:Figure_8}(b)-(c) belong to the FV regime. This distinction indicates that only when the disk stops at very small gaps does the dipole impinge directly on the bed (CV), whereas in most other cases, erosion is governed by the far-field action of the starting vortex (FV). The fits reported in Fig.~\ref{fig:Figure_8} thus successfully discriminate between the two regimes and capture the scaling of the erosion threshold with the control parameters in both cases.

Using $u_{\rm{CV,FV}}$ we can estimate the corresponding inertial Shields number $\text{Sh}_{\text{CV,FV}}=\rho u_{\text{CV,FV}}^{2}/(\Delta\rho\,g\,d)$. The critical values of these Shields numbers calculated from the critical time $\tau_c$ measured at the erosion threshold are reported in Figs.~\ref{fig:Figure_10}(a)-(b) as a function of $b/a_1$. We observe a rather good collapse of the data on the two following plateau values:
\begin{equation}
\text{Sh}_{\text{CV}}^{c}\simeq 39\pm15, 
\qquad
\text{Sh}_{\text{FV}}^{c}\simeq 22\pm12,
\label{eq:ShCVFV}
\end{equation}
for $b/a_1 \lesssim1.2$ and $b/a_1 \gtrsim1.2$ respectively. The dispersion of the data around the plateau value $\text{Sh}_{\text{CV}}^{c}\simeq 39$ in Fig.~\ref{fig:Figure_10}(a) is not so large, less than 40 \%. The larger dispersion — exceeding 50 \% of the data around the plateau value $\text{Sh}_{\text{FV}}^{c}\simeq 22$ in Fig.~\ref{fig:Figure_10}(b) — is likely due to the fact that erosion may not originate from the primary starting vortex, but rather from the secondary stopping vortex or from a tertiary vortex whose circulation and core radius may slightly differ. These values of Shields numbers are relatively high because they are not based on the friction velocity, which was not measured, but the corresponding local Shields number would certainly be roughly $10^{-2}$ smaller. This observation confirms that erosion is governed by inertial stresses in both CV and FV scenarios, but with different characteristic velocities. The particle Reynolds number ${\rm Re}_p = u_{\text{CV,FV}}d/\nu$ based on the circumferential velocity $u_{\text{CV}}$ and $u_{\text{FV}}$ exceeds always 80 throughout the explored parameter space, further supporting the inertial nature of the destabilizing force. Note that, considering the values of $u_{\text{CV}}$ and $u_{\text{FV}}$, the following scalings are found for the critical disk velocity $V_m^c = \pi L/(2 \tau_c)$ at the erosion threshold and within the explored experimental parameters $b$, $D$ and $L$:
\begin{equation}
V_{m,\text{CV}}^c \simeq \frac{\pi^2}{17} \, \frac{b^{0.14} \, L^{0.32}}{D^{0.46}} \, u_{\text{CV}}^c, 
\qquad
u_{\text{FV}} \simeq \frac{\pi^2}{1.1} \, \frac{b^{1.28}}{D^{1.05} \, L^{0.23}} \, u_{\text{FV}}^c.
\label{eq:VmCVFVscaling}
\end{equation}

The second scaling corresponding to FV explains the variations of $V_m^c$ reported in Fig.~\ref{fig:Figure_8}(a-c), where $V_m^c$ increases with $b$ but decreases for increasing $D$ and $L$ except for the first two data points of Fig.~\ref{fig:Figure_8}(a) that correspond to the CV scaling.

In summary, the erosion occurring after the disk stops is controlled by the impact of the vortices generated during the stroke. A single geometric parameter, the ratio $b/a_1$, selects whether the bed is eroded by the immediate dipole impact (CV) when the disk stop is close to the granular bed ($b/a_1 \lesssim 1$) or by the subsequent far-field action of the vortex (FV) when the disk stop is far from the granular bed ($b/a_1 \gtrsim 1$). When expressed in terms of the physically relevant local velocity, the critical Shields number is nearly constant within each regime, providing a criterion for vortex-driven grain motion that is independent of the disk kinematics.

%%%%%%%%%%%%%%%%%%%%%%%%%%%%%%%%%%%%%%%%%%%%%%%%%%%%%%%%%%%%%%
%%%%%%%%%% EROSION - MOTION AWAY FROM THE BED %%%%%%%%%%%%%%%%%%
%%%%%%%%%%%%%%%%%%%%%%%%%%%%%%%%%%%%%%%%%%%%%%%%%%%%%%%%%%%%%%

\section{Erosion by a disk moving away from a granular bed}
\label{sec:away}

%%%%%%%%%% Phenomenology %%%%%%%%%%%%%%%%%%
\subsection{Phenomenology}\label{sec:phenomenology_away}

\begin{figure}
  \centerline{\includegraphics[width=\linewidth]{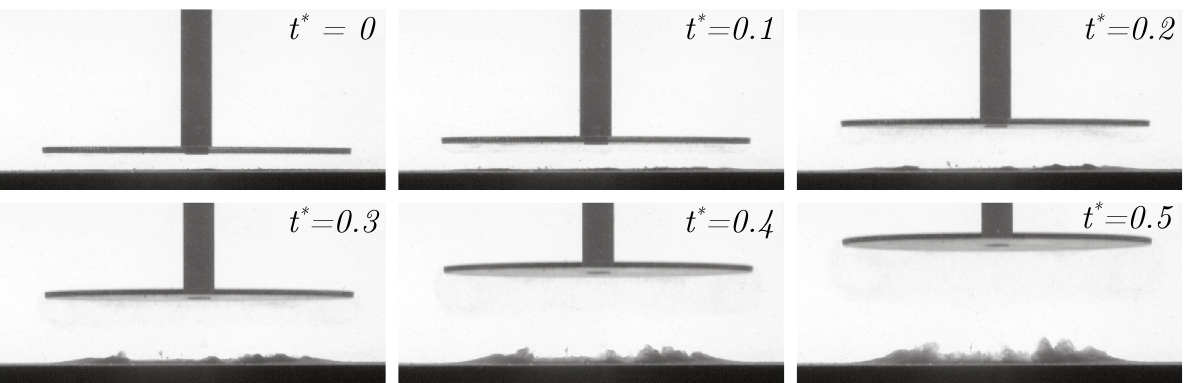}}
\caption{Time lapse showing the erosion of the granular bed produced by the unidirectional upward translation over the stroke length $L = 5.2\ \mathrm{cm}$ and time $\tau \simeq 0.4\ \mathrm{s}$ of a disk of diameter $D = 10\ \mathrm{cm}$ starting from its minimal position $b = 0.5\ \mathrm{cm}$ during the first accelerating phase ($0 \leqslant t^\ast \leqslant 0.5$) up to the disk velocity $V_m = 20$ cm/s $\simeq 2.5 V_m^c$. }
  \label{fig:Figure_11}
\end{figure}

When the disk moves away from the granular bed, a single erosion mechanism is observed, as the initial vortex ring forms beneath the translating disk and therefore never impinges directly on the bed. The erosion threshold associated with the resulting radial inflow is nevertheless modified compared with the approaching-disk case, as the confined vortex increases both radial inflow and wall shear stress.

A representative experiment is shown in Fig.~\ref{fig:Figure_11}, where the disk velocity is set well above the critical value $V_m^C$, allowing each stage to be clearly visible during the disk motion ($0 < t^\ast < 1$). At the very early time $t^{*}=0.1$, a few grains already start to lift off, forming a mound near the granular bed. As time progresses, even more grains are entrained. They move mainly upward and inward, driven by the combined action of the inward radial flow that feeds the widening gap and the starting vortex in the wake of the disk. Because this suction flow is established as soon as the vortex begins to form, erosion is observed very early in the stroke. The deformation of the bed remains almost axisymmetric and relatively shallow compared with the erosion pattern generated when the disk approaches the bed. 

%%%%%%%%%% Erosion threshold %%%%%%%%%%%%%%%%%%
\subsection{Erosion threshold} \label{sec:DPB}

\begin{figure}
  \centerline{\includegraphics[width=1.1\linewidth]{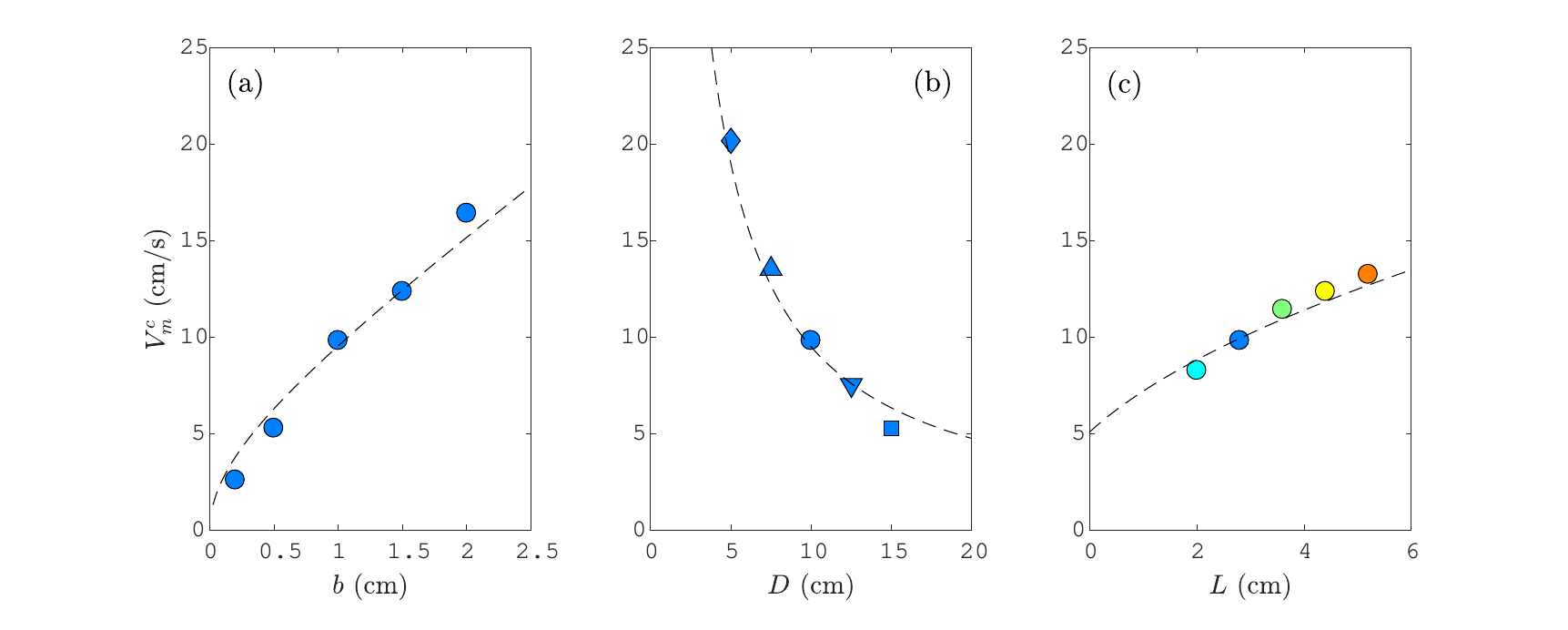}}
\caption{Critical disk velocity at the onset of erosion, $V_m^{c}$, for a disk starting from its lower position as a function of (a) the minimum distance to the granular bed, $b$ ($L = 2.8\ \mathrm{cm}$, $D = 10\ \mathrm{cm}$), (b) the disk diameter, $D$ ($L = 2.8\ \mathrm{cm}$, $b=1\ \mathrm{cm}$), and (c) the stroke length, $L$ ($D = 10\ \mathrm{cm}$, $b=1\ \mathrm{cm}$). Dashed lines are best fits from Eq.~\eqref{eq:Vmcrit} corresponding to (a) $V_m^c = 4.9 b \sqrt{1+2.8/b}$, (b) $V_m^c = 97 / D$ and (c) $V_m^c = 5.1 \sqrt{1 + L}$ with (a) $U_m^c = 12.25\ \mathrm{cm/s}$, (b) $U_m^c = 12.44\ \mathrm{cm/s}$, and (c) $U_m^c = 12.75\ \mathrm{cm/s}$.}
  \label{fig:Figure_12}
\end{figure}

The measured disk velocity $V_m^{c}$ for the onset of erosion for a disk that starts from its lower position $b$ and therefore moves away from the granular bed is reported in Figs.~\ref{fig:Figure_12}(a)-(c) when each parameter $b$, $D$, or $L$ is varied separately. We observe that $V_m^{\,c}$ (i) increases with the minimum gap distance $b$ [Fig.~\ref{fig:Figure_12}(a)], (ii) decreases with the disk diameter $D$ [Fig.~\ref{fig:Figure_12}(b)], and (iii) increases with the stroke length $L$ [Fig.~\ref{fig:Figure_12}(c)]. These trends are similar to those obtained when the disk moves towards the bed, yet the absolute values are systematically lower, showing that erosion is more easily triggered when the disk moves away from the granular bed, which may be due to the vortex formation within the gap during the disk motion that contributes to erosion. 
Dashed lines in Fig.~\ref{fig:Figure_12} show the \(V_m^{c}(b,D,L)\) best fits of the data from Eq.~\eqref{eq:Vmcrit} as the same analysis performed for the squeezing flow in the case where the disk moves towards the granular be can also be considered for the present case of suction flow where the disk moves away from the granular bed \citep{Steiner2025}. The good fit of the data with close-fitting values for $U_m^{c}$ for the corresponding set of parameters suggests that the measured erosion thresholds should obey this relationship.

\begin{figure}
  \centerline{\includegraphics[width=0.9\linewidth]{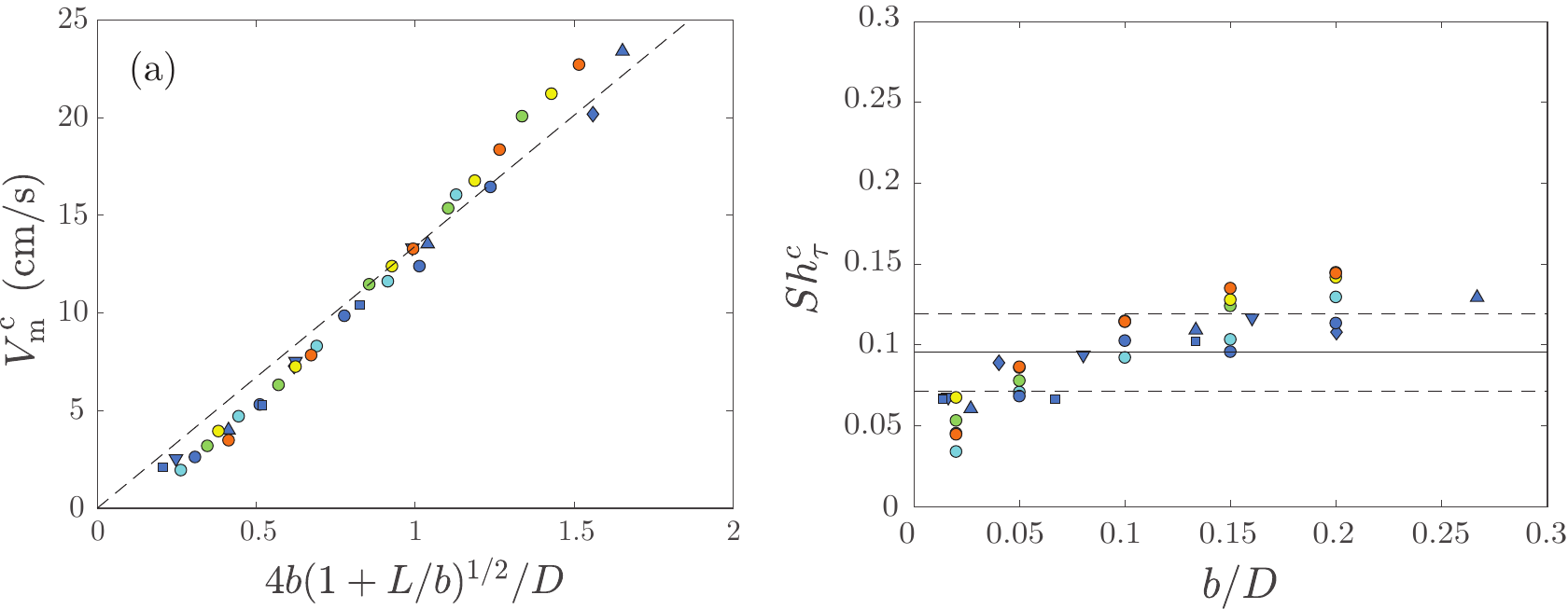}}
\caption{(a) Critical disk velocity at the onset of erosion, $V_m^{c}$, as a function of $4b\bigl(1 + L/b\bigr)^{1/2}/D$ for a disk moving away from the granular bed. The dashed line corresponds to the best fit by Eq. (3.1) with $U_m^{c} = 13.4\ \mathrm{cm\,s^{-1}}$. (b) Critical Shields number, $Sh^{c}_{\tau}$, as a function of $b/D$. The solid line is the average value $\overline{Sh^{c}_{\tau}} \simeq 0.095$, while the dotted lines delimit this mean by $\pm25\%$.}
  \label{fig:Figure_13}
\end{figure}

To test if the erosion threshold is only due to the radial suction flow, the measured disk velocity $V_m^c$ at threshold is plotted in Fig.~\ref{fig:Figure_13}(a) against the theoretical gap–averaged maximum velocity given by Eq.~\eqref{eq:Vmcrit} for all our data. All the data collapse rather well onto a master curve which is close to a linear law with a slope corresponding to $U_m^{\,c}=(13.4 \pm 1.4) \, \mathrm{cm\,s^{-1}}$, which is indeed significantly lower that the value  $U_m^{\,c}=(21.3 \pm 2.2) \, \mathrm{cm\,s^{-1}}$ found in the case where the disk moves toward the granular bed.

Following the procedure outlined in section \ref{sec:Eros_squeez}, the friction velocity $u_{\tau}$ in the viscous sublayer is also extracted from the direct numerical simulations of \cite{Steiner2025}. For each configuration we evaluate the wall shear stress $\tau_{w}= \eta \, (\partial u/\partial z)_{z=0}$ and deduce the corresponding friction velocity at the wall $u_{\tau}= (\tau_{w}/\rho)^{1/2}$. Across the full range of disk–wall separations explored, the ratio of this friction velocity to the gap-averaged velocity remains essentially constant with the value $u_{\tau}/ U_{m} \simeq 0.17 \pm 0.04$ which is larger than the value $u_{\tau}/ U_{m} \simeq 0.13 \pm 0.04$ found in the case where the disk moves toward the wall. Because the particle Reynolds number $\mathrm{Re}_p=u_\tau d/\nu$ is always larger than unity, the bed shear stress is inertial and the pertinent local Shields number is thus $\mathrm{Sh}_{\tau}=(\rho u_\tau^{2})/(\Delta\rho\,g\,d)$.
The corresponding critical value $\mathrm{Sh}^{\,c}_{\tau}$ for bed erosion threshold is plotted as a function of the relative minimum distance of the disk to the granular bed $b/D$ in Fig.~\ref{fig:Figure_13}(b). The mean value for this critical Shields number is  $\overline{\mathrm{Sh}^{\,c}_{\tau}}\simeq0.095 \pm 0.01$. When compared to the case where the disk moves toward the granular bed, this mean value is significantly smaller, and there is also a significant trend of smaller $\mathrm{Sh}^{\,c}_{\tau}$ for smaller $b/D,$ which can be observed despite the scattering of the data. This value is also smaller than the value $\rm{Sh}_{\tau}^c = 0.13 \pm 0.01$ known for steady homogeneous flows \citep{Loiseleux2005,Ouriemi2007}. This likely reflects that the model neglects the influence of the vortex on the gap-driven inward radial flow.

\begin{figure}
  \centerline{\includegraphics[width=0.9\linewidth]{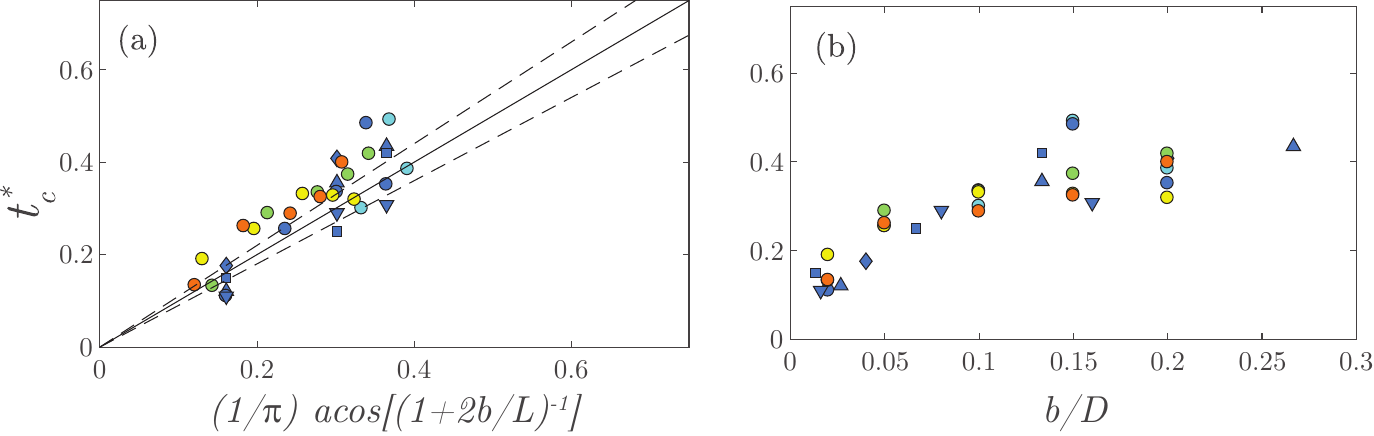}}
\caption{Dimensionless time $t_c^\ast$ at which the first grains begin to move at the erosion threshold, with the disk released from its lower position, as a function of (a) $t_m^\ast=(1/\pi){\rm acos}{(1+2b/L)^{-1}}$ and (b) $b/D$. }
  \label{fig:Figure_14}
\end{figure}

To investigate further the erosion process, let us now look at the dimensionless time $t_\mathrm{c}^\ast$ at which the first grains are observed to move just above the threshold, which is shown in Fig.~\ref{fig:Figure_14}. When the disk moves away from the granular bed, erosion always starts early, $0.1 \lesssim t_\mathrm{c}^\ast\lesssim 0.4$, \textit{i.e.}, before the disk reaches its maximum velocity at $t^\ast = 0.5$. As observed in Fig.~\ref{fig:Figure_14}(a) the measured values of $t_\mathrm{c}^\ast$ are highly scattered around the theoretical time $t_m^\ast=(1/\pi){\rm acos}(1+2b/L)^{-1}$ where the radial velocity of the suction flow is expected to be maximal from mass conservation \citep{Steiner2025}. This suggests that erosion is not solely caused by the gap-averaged radial flow and that vortex formation is responsible for this data dispersion. In Fig.~\ref{fig:Figure_14}(b), it can be seen that the onset time of erosion increases significantly with $b/D$. Depending on the instant when erosion begins, the vortex will be more or less developed and more or less distant from the granular bed. 

Figures~\ref{fig:Figure_13}(b) and~\ref{fig:Figure_14}(b) show that both the critical Shields number $\mathrm{Sh}^{\,c}_{\tau}$ and the critical time $t_{\mathrm{c}}^{\ast}$ for erosion increase with $b/D$. Indeed, at small gaps ($b/D\lesssim0.2$) the vortex core lies only a few grain diameters above the bed and the circumferential velocity decay as $1/r'$ increases the gap–driven inflow, so grains lift off almost instantly. As the gap widens, the core moves away and its contribution to the near-wall shear decays roughly as $1/b$. A larger purely gap-driven inflow is therefore needed to reach the same bed stress, which delays erosion (larger $t_{\mathrm{c}}^{\ast}$) and raises the value of the critical Shields number $Sh^{c}_{\tau}\propto {U_{m}^c}^{2}$.

In summary, when the disk moves away from the granular bed, the erosion is triggered during the first accelerating phase of the disk motion by the radial inflow generated between the disk and the granular surface, which is reinforced by the starting vortex ring. The threshold velocity follows the scaling law of Eq.~\eqref{eq:Vmcrit}, with a prefactor approximately $40\%$ smaller than in the approaching-disk case.

%%%%%%%%%%%%%%%%%%%%%%%%%%%%%%%%%%%%%%%%%%%%%%%%%%%%%%%%%%%%%%
%%%%%%%%%%%%%%%%%%%%%%%% CONCLUSION %%%%%%%%%%%%%%%%%%%%%%%%%%
%%%%%%%%%%%%%%%%%%%%%%%%%%%%%%%%%%%%%%%%%%%%%%%%%%%%%%%%%%%%%%

\section{Conclusion}
\label{sec:conclusion}

In this study, we experimentally investigated how a rigid disk of diameter $D$ translating towards or away from a granular bed can erode and ultimately resuspend glass beads. We have focused here on an elementary unsteady motion at the minimal distance $b$ from the granular bed: a single stroke of length $L$ during the time $\tau$ at the maximum velocity $V_m$, and have quantified the onset of erosion $V_m^c$ and the dimensionless time $t_c^\ast = t_c/\tau$ at which it arises.

When the disk moves towards the bed, we found that the initial erosion is triggered by the radial outflow that develops beneath the disk while it is still decelerating ($0.5 < t_c^\ast < 1$). For sufficiently small gaps ($b/D\lesssim0.2$), the threshold is governed by a constant critical outflow velocity $U_m^c \simeq V_m^c D\!\left[4b(1+L/b)^{1/2}\right]^{-1}$ that is related to the disk motion close to the granular bed from mass conservation. The critical local Shields number based on the friction velocity $u_\tau \simeq 0.13 U_m$ at the bed level is $\text{Sh}_\tau^c \simeq 0.19$. Once the disk stops, the interaction between the starting vortex ring and the edge of the disk gives rise to a dipole of vortex rings. This dipole produces either immediate ($t_c^\ast \simeq 1$) or delayed ($t_c^\ast > 1$) erosion depending on whether the dipole is close (CV) or far (FV) from the bed, respectively, {\it i.e.} whether the ratio $b/a_1$ is smaller or larger than unity, respectively, where $a_1$ is the radius of the vortex core at the disk stop. This complex vortex flow explains both the bidirectional grain transport observed near the bed and the increase in the disk critical velocity with $b/a_1$.

In the opposite case, when the disk translates away from the granular bed, erosion occurs before the disk reaches its maximum velocity ($0 < t_c^\ast < 0.5$), i.e., during its first accelerating phase, due to an inward radial suction reinforced by the developing starting vortex within the gap. This is the reason why the corresponding critical Shields number $\text{Sh}_\tau^c \simeq 0.095$ is significantly smaller than the value 0.19 corresponding to the disk motion towards the granular bed and decreases for decreasing $b/D$.

Taken together, the two configurations highlight how asymmetries in the flow, depending on the direction of motion (outflow versus inflow, starting versus stopping vortices), result in different erosion thresholds. The asymmetry in the flow generated also explains why a disk that travels away from the granular bed, despite generating a weaker wall stress, can still mobilize grains earlier in the stroke owing to the rapid growth of the starting ring. These findings provide a quantitative framework for guiding geophysical or biological scenarios in which impulsively moving objects induce sediment transport via vortex shedding. An open issue is how the erosion and transport threshold is altered when the disk (or a biological analogue such as a flounder fin) performs repeated oscillations rather than a single stroke, because the vortices shed during successive half-cycles are expected to interact with one another in a more complex, history-dependent manner. Erosion has already yielded some surprising results for vertical oscillations of a horizontal disk \citep{LaRagione2019} or the horizontal oscillations of a vertical plate \citep{Martino2018} near a granular bed.

%\begin{bmhead}[Xxxxxxx.]
%For the custom heading, such as acknowledgment, funding disclosure,
%conflict disclosure and any other like-wise sections must be
%mentioned in the optional braces as shown in this example.
%\end{bmhead}

% \backsection[Supplementary data]{\label{SupMat}Supplementary material and movies are available at \\https://doi.org/10.1017/jfm.2019...}
%
 \backsection[Acknowledgements]{The authors thank  Ivan Delbende for his help in the numerical simulations, Guillaume Quibeuf and Guillaume Renard for their preliminary experimental contributions, and Johannes Amarni, Alban Aubertin, Lionel Auffray, and Rafael Pidoux for their work on the experimental setup.}
%
% \backsection[Funding]{}
%
 \backsection[Declaration of interests]{The authors report no conflict of interest.}
%
% \backsection[Data availability statement]{The data that support the findings of this study are openly available in [repository name] at http://doi.org/[doi], reference number [reference number]. See JFM's \href{https://www.cambridge.org/core/journals/journal-of-fluid-mechanics/information/journal-policies/research-transparency}{research transparency policy} for more information}
%
\backsection[Author ORCIDs]{Joanne Steiner, \url{https://orcid.org/0009-0007-6747-594X};
Cyprien Morize, \url{https://orcid.org/0000-0002-6966-648X};
Alban Sauret, \url{https://orcid.org/0000-0001-7874-5983};
Philippe Gondret, \url{https://orcid.org/0000-0002-7184-9429}.}

\appendix
\section{Scalings of the starting and stopping vortices in the wake of a disk}
\label{appendix:Unbounded}

\subsection{Vortex scalings in the unbounded case}

In the unbounded case, where the disk of diameter $D$ is far from any boundaries ($b/a \gg 1)$, the {\it maximum} circulation $|\Gamma_m|$ and core radius $a_m$ of the {\it starting} vortex ring formed in the wake of a translating disk whose velocity evolves sinusoidally over the stroke length $L$ and time $\tau$  have been shown by \cite{Steiner2023} to follow the scaling laws
\begin{equation}
|\Gamma_m| \simeq 2.1 \, L^{4/3}D^{2/3} \tau ^{-1}  \quad \mathrm{and} \quad  a_m \simeq 0.1 \, L^{2/3}D^{1/3}.
\end{equation}

The {\it maximum} circulation and core radius of the {\it starting} vortex have been reported by \cite{Steiner2023} to be reached at the dimensionless times $\tau_{\Gamma_m}^\ast = 0.70 \pm 0.02$ and $\tau_{a_m}^\ast = 0.79 \pm 0.07$, respectively, which do not depend on the size ratio $L/D$. \cite{Steiner2023} have also shown that the circulation and core radius of the {\it starting} vortex at the {\it disk stop} ($t^\ast = 1$) follow the same scaling laws with only slightly smaller numerical coefficients:

\begin{equation}
|\Gamma_1| \simeq 1.9 \, L^{4/3}D^{2/3} \tau ^{-1}  \quad \mathrm{and} \quad  a_1 \simeq 0.085 \, L^{2/3}D^{1/3}.
\end{equation}

And finally, \cite{Steiner2023} have also reported that the scalings of the {\it maximum} circulation and core radius of the counter-rotating {\it stopping} vortex are the same, with even smaller coefficients:
\begin{equation}
|\Gamma_{m,s}| \simeq 1.1 \, L^{4/3}D^{2/3} \tau ^{-1}  \quad \mathrm{and} \quad  a_{m,s} \simeq 0.06 \, L^{2/3}D^{1/3}.
\end{equation}

The {\it maximum} circulation and maximum core radius of the {\it stopping} vortex are thus about half the maximum values measured for the primary {\it starting} vortex. They are reached at the dimensionless times $\tau_{\Gamma_{m,s}}^\ast= 1.23 \pm 0.05$ and $\tau_{\Gamma_{m,s}}^\ast= 1.27 \pm 0.09$, respectively.

\subsection{Vortex scalings close to a wall}

In the presence of a bottom wall, the minimum distance $b$ between the disk and the wall has been shown to play a significant role in the generation process of the vortex ring \cite[][]{Steiner2025}. The dimensionless coefficients for the circulation $c_\Gamma$ and the core radius $c_a$ are defined as
\begin{equation}
c_\Gamma = \frac{|\Gamma|}{L^{4/3}D^{2/3} \tau ^{-1}}, \quad \mathrm{and} \quad c_a = \frac{a}{L^{2/3}D^{1/3}},
\label{eq:AdimCoefficients}
\end{equation}
have been shown to depend on the two ratios $b/D$ and $L/D$ for close-wall motions, rather than being constant for far-wall motions, and to differ between the approaching and leaving cases, as detailed in the following.

\subsubsection{Vortex scalings in the approaching case}
\label{appendix:A.2.1}
When the disk is translating {\it toward} the wall, the dimensionless coefficients for the {\it maximum} circulation $c_{\Gamma_m}$ and the maximum radius $c_{a_m}$ of the vortex core have been shown by \cite{Steiner2025} to follow the scalings

\begin{equation}
c_{\Gamma_m} \simeq \left\{ 
\begin{array}{l}
2.1 \\
\\
1.3 \, (L/D)^{-0.1}(b/D)^{-0.28}
\end{array}
\quad \mathrm{for} \quad (L/D)^{-0.1}(b/D)^{-0.28} \, \left\{
\begin{array}{l}
\lesssim 1.6 \\
\\
\gtrsim 1.6
\end{array}
\right.
\right.
\end{equation}

\begin{equation}
c_{a_m} \simeq \left\{ 
\begin{array}{l}
0.1 \\
\\
6.8 \times10^{-2} (L/D)^{-0.12}(b/D)^{-0.14}
\end{array}
\quad \mathrm{for} \quad (L/D)^{-0.12}(b/D)^{-0.14} \, \left\{
\begin{array}{l}
\lesssim 1.4 \\
\\
\gtrsim 1.4
\end{array}
\right.
\right.
\end{equation}

The {\it maximum} circulation $\Gamma_m$ and core radius $a_m$ of the {\it starting} vortex in the near-wall configuration for the {\it approaching case} can be rewritten in dimensional form as

\begin{equation}
  |\Gamma_m| \simeq 1.3 \, L^{1.23}D^{1.05}b^{-0.28}\tau^{-1}\quad \mathrm{and} \quad a_m \simeq 6.8\times 10^{-2}L^{0.55}D^{0.59}b^{-0.14}.
\end{equation}

\begin{figure}
  \centerline{\includegraphics[width=\linewidth]{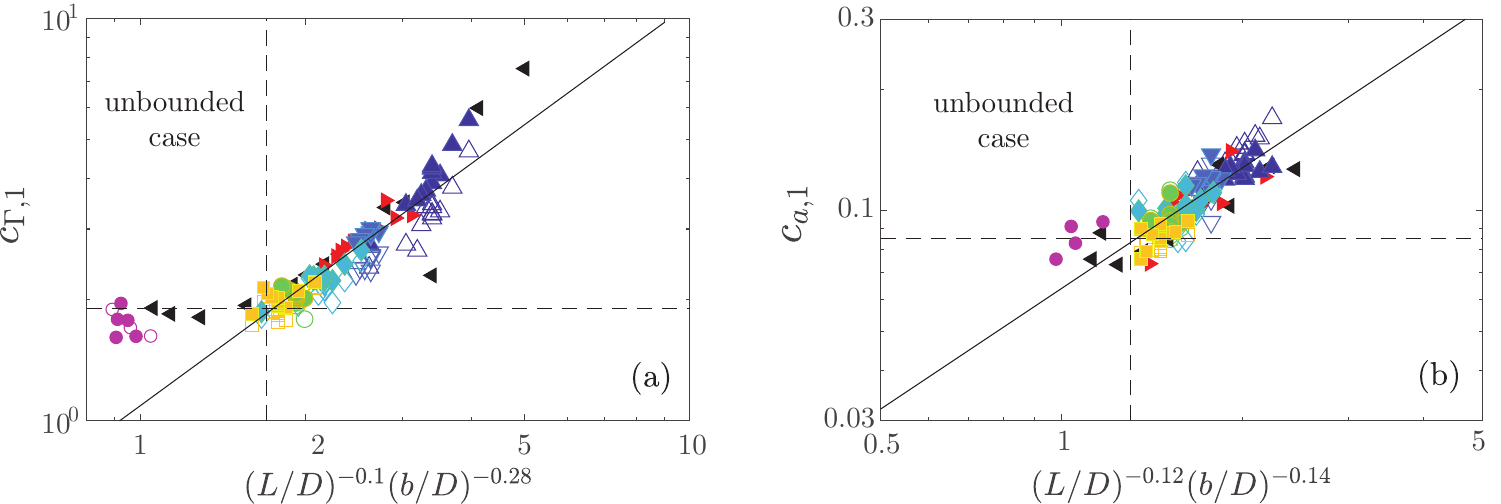}}
\caption{Dimensionless (a) circulation $c_{\Gamma_1}$ as a function of $(L/D)^{-0.1}(b/D)^{-0.28}$, and (b) dimensionless core radius $c_{a_1}$ as a function of $(L/D)^{-0.12}(b/D)^{-0.14}$ for a disk moving toward a wall. The horizontal dashed lines are $c_{\Gamma_1, \infty} \simeq 1.9$ and $c_{a_1, \infty} \simeq 0.085$, which correspond to the values obtained in an infinite medium by \cite{Steiner2023}. The equations of the solid lines are (a) $c_{\Gamma_1} = 1.1 (b/D)^{-0.28}(L/D)^{-0.1}$ and (b) $c_{a_1} = 0.064 (b/D)^{-0.14}(L/D)^{-0.12}$. The vertical dashed lines are $(L/D)^{-0.1}(b/D)^{-0.28} = 1.7 \, \text{(a)}$ and $1.3 \, \text{(b)}$. All the open (resp. filled) symbols correspond to the experiments  (resp. simulations) reported in Table~\ref{tab:Symbols}.}
  \label{fig:Figure_15}
\end{figure}

The circulation $\Gamma_1$ and core radius $a_1$ of the {\it starting} vortex ring at the disk stop ($t^\ast = 1$) have not been reported by \cite{Steiner2025}, but \cite{Steiner2023PhD} has shown that they follow the same scalings as the maximum values with only slightly smaller coefficients:

\begin{equation}
c_{\Gamma_1} \simeq \left\{ 
\begin{array}{l}
1.9 \\
\\
1.1 \, (L/D)^{-0.1}(b/D)^{-0.28}
\end{array}
\quad \mathrm{for} \quad (L/D)^{-0.1}(b/D)^{-0.28} \, \left\{
\begin{array}{l}
\lesssim 1.7 \\
\\
\gtrsim 1.7
\end{array}
\right.
\right.
\end{equation}

\begin{equation}
c_{a_1} \simeq \left\{ 
\begin{array}{l}
0.085 \\
\\
6.4 \times10^{-2} (L/D)^{-0.12}(b/D)^{-0.14}
\end{array}
\quad \mathrm{for} \quad (L/D)^{-0.12}(b/D)^{-0.14} \, \left\{
\begin{array}{l}
\lesssim 1.3 \\
\\
\gtrsim 1.3
\end{array}
\right.
\right.
\end{equation}

The measured values for the dimensionless coefficients $c_{\Gamma_1}$ and $c_{a_1}$ from the PIV experimental measurements (open symbols) and numerical simulations (filled symbols) described in \cite{Steiner2023PhD} are reported in Fig.~\ref{fig:Figure_15} for the flow configurations reported in Table~\ref{tab:Symbols}.
The circulation $\Gamma_1$ and radius $a_1$ of the {\it starting} vortex at the {\it disk stop} ($t^\ast = 1$) in the near-wall configuration for the {\it approaching} case can be rewritten in dimensional form as

\begin{table}
\centering
\setlength{\tabcolsep}{6pt}
\begin{tabular}{ c c c c c c c c } 
 \hline  \hline 
 $b$ (cm) & $L$ (cm) & $D$ (cm) & $\tau$ (s) & $L/D$ & $b/D$ & $Re$ & Symbols\\
 \hline
0.2 & 2 - 5.2 & 5 - 15 & 0.5 - 0.52 & 0.19 - 0.56 & 0.013 - 0.04 & $1.3\times10^3$ - $8.8\times10^3$ &  \color{blue}{$\triangle$}/ \color{blue}{$\blacktriangle$}  \\
0.5 & 2 - 5.2 & 10 & 0.5 - 2.5  & 0.2 - 0.52 & 0.05 & $1.8\times10^3$ - $8.8\times10^3$  & \color{blue}{$\triangledown$} / \color{blue}{$\blacktriangledown$} \\
1 & 2 - 5.2 & 5 - 15 & 0.5 - 2.5  & 0.19 - 0.56 & 0.067 - 0.2 & $1.8\times10^3$ - $9.8\times10^3$ &  \color{blue}{$\blacklozenge$} / \color{blue}{$\blacklozenge$} \\
1.5 & 2 - 5.2 & 10 & 0.5 - 2.5  & 0.2 - 0.52 & 0.15 & $1.8\times10^3$ - $1.3\times10^4$ &  \color{green}{$\circ$} / \color{green}{$\bullet$} \\
2 & 2 - 5.2 & 7.5 - 15 & 0.36 - 0.83  & 0.19 - 0.52 & 0.13 - 0.27 & $5.3\times10^3$ - $2\times10^4$ &  \color{yellow}{$\square$} / \color{yellow}{$\blacksquare$} \\
 0.2 - 13 & 2.8 - 11.2 & 10 - 40 & 0.5  & 0.28 & 0.005 - 1.3 & $1.5\times10^4$ &  $\blacktriangleleft$ \\
0.4 - 2 & 2 - 16 & 8-40 & 0.5 & 0.05 - 2 & 0.05 & $1.5\times10^4$&  \color{red}{$\blacktriangleright$} \\
20 & 2 - 20 & 3.75-40 & 0.25-2.5 & 0.07 - 2 & 0.5-5.3 & $10^3$ - $3\times10^4$&  \color{violet}{$\circ$} / \color{violet}{$\bullet$} \\
 \hline  \hline 
\end{tabular}
 \caption{Sets of experimental and numerical parameters, associated non-dimensional numbers, and corresponding symbols used for Fig.~\ref{fig:Figure_15}. Empty symbols correspond to experiments, and full symbols correspond to numerical simulations.}
 \label{tab:Symbols}
\end{table}

\begin{equation}
  |\Gamma_1| \simeq 1.1 \, L^{1.23}D^{1.05}b^{-0.28}\tau^{-1}\quad \mathrm{and} \quad a_1 \simeq 6.4\times 10^{-2}L^{0.55}D^{0.59}b^{-0.14}.
  \label{eq:circulation_radius_toward}
\end{equation}

\subsubsection{Vortex scalings in the leaving case}

When the disk is translating {\it away} from the wall, the dimensionless coefficient for the {\it maximum} circulation $c_{\Gamma_m}$ of the {\it starting} vortex has been shown by \cite{Steiner2025} to follow the scalings

\begin{equation}
c_{\Gamma,m} = \left\{ 
\begin{array}{l}
2.1 \\
\\
1.13 \, (L/D)^{-0.18}(b/D)^{-0.28}
\end{array}
\quad \mathrm{for} \quad (L/D)^{-0.18}(b/D)^{-0.28} \,\left\{
\begin{array}{l}
< 1.9 \\
\\
> 1.9
\end{array}
\right.
\right.
\label{eq:adim_circulation_away}
\end{equation}

The {\it maximum} circulation $\Gamma_m$ of the {\it starting} vortex in the near-wall configuration for the {\it leaving} case can be rewritten in dimensional form as

\begin{equation}
    |\Gamma_m| = 1.13 \, L^{1.15}D^{1.12}b^{-0.28}\tau^{-1}.
\end{equation}

The {\it maximum} core radius of the {\it starting} vortex for the leaving case has been shown by \cite{Steiner2025} to not be changed from the unbounded case.

\bibliographystyle{jfm}
\bibliography{jfm}

\end{document}